\documentclass[aps,twocolumn,superscriptaddress,footinbib,longbibliography]{revtex4-1} 

\usepackage{longtable}
\usepackage{morefloats}
\usepackage[dvips]{graphicx}
\usepackage{xcolor}
\usepackage{epsfig,graphicx,amsfonts,amsbsy}
\usepackage{amsmath,amsfonts,amsthm,amssymb}
\usepackage{appendix}
\usepackage{makeidx}
\usepackage{url}
\usepackage{verbatim}
\usepackage[bookmarksnumbered,pdfpagelabels=true,plainpages=false,colorlinks=true,linkcolor=blue,citecolor=red,urlcolor=blue]{hyperref}
\usepackage[rightcaption]{sidecap}
\usepackage{array}
\usepackage{multirow}
\usepackage{tabularx}
\usepackage{braket} 
\usepackage{dsfont}
\usepackage{bbold}
\usepackage{etoolbox}
\usepackage{units}
\usepackage{filecontents}

\usepackage{blindtext}
\usepackage{comment}
\usepackage{amssymb}

\def\vec{{\vec{S}}}

\def\CalE{{\mathcal{E}}}
\def\CalM{{\mathcal{M}}}
\def\CalP{{\mathcal{P}}}
\def\CalB{{\mathcal{B}}}
\def\CalQ{{\mathcal{Q}}}

\usepackage{newtxtext}
\usepackage{newtxmath}
\renewcommand{\vec}[1]{\ensuremath{\boldsymbol{#1}}}
\newcommand\titlelowercase[1]{\texorpdfstring{\lowercase{#1}}{#1}}

\AtBeginDocument{%
    \newwrite\bibnotes
    \def\bibnotesext{Notes.bib}
    \immediate\openout\bibnotes=\jobname\bibnotesext
    \immediate\write\bibnotes{@CONTROL{REVTEX41Control}}
    \immediate\write\bibnotes{@CONTROL{%
    apsrev41Control,author="08",editor="1",pages="1",title="0",year="1"}}
     \if@filesw
     \immediate\write\@auxout{\string\citation{apsrev41Control}}%
    \fi
}%

\begin{document}

\title{
Magnetoelectric Cavity Magnonics in Skyrmion Crystals
}

\author{Tomoki Hirosawa}
\thanks{These two authors contributed equally.}
\affiliation{Department of Physics, University of Basel, Klingelbergstrasse 82, CH-4056 Basel, Switzerland}
\affiliation{Department of Physical Sciences, Aoyama Gakuin University, Sagamihara, Kanagawa 252-5258, Japan}

\author{Alexander Mook}
\thanks{These two authors contributed equally.}
\affiliation{Department of Physics, University of Basel, Klingelbergstrasse 82, CH-4056 Basel, Switzerland}
\affiliation{Department of Physics TQM, Technical University of Munich, James-Franck-Stra{\ss}e 1, D-85748 Garching, Germany}
\affiliation{Institute of Physics, Johannes Gutenberg University Mainz, D-55128 Mainz, Germany}

\author {Jelena Klinovaja}
\affiliation{Department of Physics, University of Basel, Klingelbergstrasse 82, CH-4056 Basel, Switzerland}

\author{Daniel Loss}
\affiliation{Department of Physics, University of Basel, Klingelbergstrasse 82, CH-4056 Basel, Switzerland}

\begin{abstract}
    We present a theory of magnetoelectric magnon-photon coupling in cavities hosting noncentrosymmetric magnets. Analogously to nonreciprocal phenomena in multiferroics, the magnetoelectric coupling is time-reversal and inversion asymmetric. This asymmetry establishes a means for exceptional tunability of magnon-photon coupling, which can be switched on and off by reversing the magnetization direction.
    Taking the multiferroic skyrmion-host Cu$_2$OSeO$_3$ with ultralow magnetic damping as an example, we reveal the electrical activity of skyrmion eigenmodes and propose it for magnon-photon splitting of ``magnetically dark'' elliptic modes.
    Furthermore, we predict a cavity-induced magnon-magnon coupling between magnetoelectrically active skyrmion excitations. We discuss applications in quantum information processing by proposing protocols for all-electrical magnon-mediated photon quantum gates, and a photon-mediated SPLIT operation of magnons.
    Our study highlights magnetoelectric cavity magnonics as a novel platform for realizing quantum-hybrid systems and the coherent transduction between photons and magnons in topological magnetic textures.
\end{abstract}
\maketitle

\section{Introduction}
In electromagnetic cavities, light-matter interactions are strongly enhanced and can bring hybrids of photons and the material's quasiparticles into being \cite{kockum2019ultrastrong}. Examples are magnon-photon (MP) hybrids in magnetic materials \cite{soykalStrongFieldInteractions2010,Huebl2013, Tabuchi2014, Goryachev2014, Zhang2014, Hou2019, Li2019}. 
Since the frequency of magnon modes can be controlled with external magnetic fields, it enables exceptionally tunable MP coupling. Furthermore, magnons provide interactions with different quantum systems such as phonons, microwave photons, and optical photons, rendering MP hybrids a promising platform for novel quantum technologies \cite{LachanceQuirion2019,harderCoherentDissipativeCavity2021a,ZARERAMESHTI20221}.
Anticipated applications include, e.g., magnon-qubit coupling \cite{Trifunovic2013,Tabuchi2015}, transducing microwave to optical quantum information \cite{Hisatomim2016}, and quantum-enhanced sensing \cite{LachanceQuirion2017sensing}.

Microscopically, several origins of MP interactions have been established, e.g., the linear coupling between microwave magnetic fields and the material's magnetic dipoles \cite{Huebl2013, Tabuchi2014, Goryachev2014, Zhang2014} and, at optical frequencies, the nonlinear coupling to electric fields \cite{Hisatomim2016, Osada2016, Zhang2016opto, Liu2016Optomag,violakusminskiyCoupledSpinlightDynamics2016}. 
Recent years have seen a growing interest in magnetoelectric (ME) materials that host electrically controllable, topologically nontrivial magnetic textures such as skyrmions \cite{nagaosaTopologicalPropertiesDynamics2013,sekiObservationSkyrmionsMultiferroic2012, sekiMagnetoelectricNatureSkyrmions2012, kezsmarki2015neel, ruff2015magnetoelectric, ruff2015multiferroicity, Fujima2017lacunar, White2018, yao2020controlling, ba2021electric,hirosawaLasercontrolledRealReciprocalspace2021}. These are well-localized magnetic defects whose bound magnons~\cite{linInternalModesSkyrmion2014,Schutte2014,Diaz2020FM,diazSpinWaveRadiation2020a} may get hybridized with superconducting qubits via cavity-mediated interactions. Alternatively, skyrmions themselves may serve as qubits \cite{Psaroudaki2021}, with cavity MP coupling realizing qubit-qubit coupling. To explore such possibilities, a detailed understanding of MP coupling in ME materials is crucial. Existing theories of cavity magnonics fall short because they do not account for 
electrically active magnons (electromagnons) that were separately studied, e.g., in Refs.~[\onlinecite{pimenovPossibleEvidenceElectromagnons2006,valdesaguilarOriginElectromagnonExcitations2009,takahashiMagnetoelectricResonanceElectromagnons2012,okamura2013microwave, kubackaLargeAmplitudeSpinDynamics2014,tokuraMultiferroicsSpinOrigin2014a, mochizukiDynamicalMagnetoelectricPhenomena2015, Okamura2015}]. 

\begin{figure}
    \centering
    \includegraphics[width=1\columnwidth]{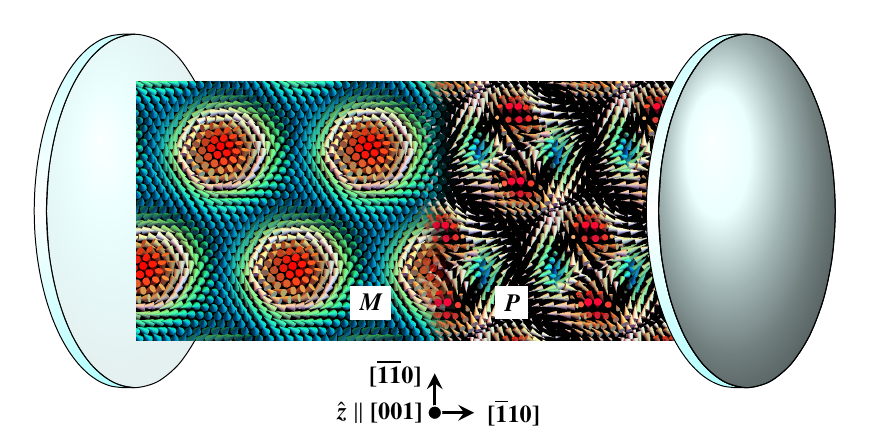}
    \caption{
        Setup for studying magnetoelectrical magnon-photon hybridization in magnets with topologically nontrivial textures. A multiferroic skyrmion crystal phase of Cu$_2$OSeO$_3$ is placed in an electromagnetic cavity. Left: Magnetization $\vec{M}$ exhibiting a Bloch skyrmion winding. Right: Polarization $\vec{P}$ with a quadrupolar texture. Both dipoles couple to the electromagnetic cavity fields, leading to a finite magnon-photon splitting. The coordinate system indicates crystallographic axes and $\hat{\vec{z}}$ the direction of a static magnetic field stabilizing the skyrmion crystal phase.
    }
    \label{fig:teaser}
\end{figure}

Herein, we extend the microscopic theory of cavity magnonics to ME materials with topologically nontrivial spin textures (see Fig.~\ref{fig:teaser}). We reveal that the resulting MP coupling is strongly anisotropic with respect to magnetization reversal, establishing a means for exceptional tunability. We make quantitative predictions for the multiferroic skyrmion host Cu$_2$OSeO$_3$ that recently drew a lot of interest in the context of cavity magnonics \cite{Abdurakhimov2019,khanCouplingMicrowavePhotons2021,liensbergerTunableCooperativityCoupled2021a}. In the ferromagnetic phase of Cu$_2$OSeO$_3$, the MP anisotropy can be so strong that MP coupling is absent for one magnetization direction but leading to strong coupling for the opposite direction. The strong coupling limit can even be reached fully electrically when magnetic dipole coupling is inactive. In the skyrmion crystal (SkX) phase of Cu$_2$OSeO$_3$, the static magnetization forms regular arrays of Bloch skyrmions~\cite{sekiObservationSkyrmionsMultiferroic2012} and the corresponding static electric polarization exhibits a quadrupole moment~\cite{sekiMagnetoelectricNatureSkyrmions2012,belesiMagnetoelectricEffectsSingle2012,whiteElectricFieldControl2012}, as shown in Fig.~\ref{fig:teaser}. The distinct symmetries of magnetization and polarization lead to an electric activity of magnons \cite{okamura2013microwave,mochizukiDynamicalMagnetoelectricPhenomena2015, Okamura2015} beyond their well-known magnetic activity~\cite{mochizukiSpinWaveModesTheir2012}. 
In a SkX, the skyrmion-skyrmion interaction allows hybridization among single-skyrmion bound states and thus results in a rich magnetoelectrical activity of skyrmion excitations. 
We predict the possibility (i) of a purely electric MP coupling between cavity and magnon modes, an example being the ``magnetically dark'' elliptical SkX mode, and (ii) of cavity-induced magnon-magnon coupling between the breathing and counterclockwise modes. 

As an application in quantum information, we demonstrate all-electrical quantum gates between two microwave photon modes coupled to the magnet, with ``all-electrical'' referring to both the magnon-frequency tuning and MP interactions being the result of electric dipole coupling. 
In particular, we propose magnon-mediated SPLIT and SWAP operations between two photons in the collinear ferromagnetic phase, and a photon-mediated SPLIT operation of magnons in the SkX phase.
We conclude that magnetoelectric cavity magnonics is a promising platform for realizing quantum-hybrid systems and quantum transduction of microwave photons and magnons, which may prove instrumental for realizing skyrmionic quantum computing and manipulating quantum states of magnons~\cite{yuanQuantumMagnonicsWhen2022}. 


The structure of this paper is as follows. In Sec.~\ref{sec: ME_theory}, we develop the theory of ME MP coupling. The bilinear ME MP coupling is obtained by expanding the coupling between electromagnetic fields and ME materials in terms of magnon operators. We also discuss the symmetry of ME MP coupling. In Sec.~\ref{sec: worked_example}, the multiferroic material Cu$_2$OSeO$_3$ is considered as a worked example. We estimate the magnetic and electric MP coupling strengths in the collinear ferromagnetic phase and the SkX phase of Cu$_2$OSeO$_3$. We show that the interplay between magnetic and electric couplings result in the asymmetry of MP coupling with respect to the magnetization direction. We also reveal the electrical activity of the elliptic mode and the cavity-mediated magnon-magnon coupling in SkXs. In Sec.~\ref{sec: QI_application}, we discuss the quantum information application of ME MP coupling. We propose protocols for all-electrical magnon-mediated quantum gates of photons in the collinear ferromagnetic phase of Cu$_2$OSeO$_3$. We also demonstrate a coherent quantum operation of magnons mediated by photons in the SkX phase of Cu$_2$OSeO$_3$. In Sec.~\ref{sec: Conclusion}, we provide a summary of this work. Appendices~\ref{sec:multipole_magnon}-\ref{sec: SkX_Stark} contain additional information and technical details.

\section{Theory of magnetoelectric magnon-photon coupling}
\label{sec: ME_theory}
We consider a ME insulator in a microwave cavity, a situation described by the Hamiltonian $\mathcal{H} = \mathcal{H}_\text{SL}+\mathcal{H}_\text{CA}+\mathcal{H}_\text{EM}$. Here, $\mathcal{H}_\text{SL}$ is a spin-lattice Hamiltonian, $\mathcal{H}_\text{CA}$ describes cavity photons, and $\mathcal{H}_\text{EM}$ contains the coupling between electric, $\vec{E}(\vec{r},t)$, and magnetic fields, $\vec{B}(\vec{r},t)$, and the material's electric polarization $\vec{P}$ and magnetization $\vec{M}$, respectively. The macroscopic moments may be expressed as a sum over lattice sites at position $\vec{r}_i$, where $i = 1,\ldots, N$, with $N$ being the total number of sites. Each site features an electric, $\vec{p}_i$, and magnetic moment, $\vec{m}_i = - \mathsf{g} \mu_\text{B} \vec{S}_i$, with Land\'e factor $\mathsf{g}$, Bohr magneton $\mu_\text{B}$, and (dimensionless) spin moment $\vec{S}_i$. At the (classical) Hamiltonian level, the electromagnetic interactions are accounted for by the Stark and Zeeman energy,
\begin{align}
    \mathcal{H}_\text{EM} 
    = 
    -\sum_{i=1}^N [ 
    \vec{E}(\vec{r}_i,t) \cdot \vec{p}_i (t)
    +
    \vec{B}(\vec{r}_i,t) \cdot \vec{m}_i (t) 
    ].
\end{align}

\subsection{Derivation of bilinear magnetoelectric magnon-photon coupling}
Hybridizations between magnetic excitations and electromagnetic waves are described by a bilinear Hamiltonian in the second-quantized language. First, we quantize the excitations in the material by performing an expansion about the magnetically ordered ground state in terms of magnon annihiliation (and creation) operators $b^{(\dagger)}_i$ \cite{Holstein1940}. To linear order in magnon operators, we find 
\begin{align}
    \vec{S}_i \approx S \hat{\vec{e}}^z_i + \sqrt{S/2} ( \hat{\vec{e}}^-_i b_i + \hat{\vec{e}}^+_i b_i^\dagger),
\end{align} 
where $S$ is the spin quantum number, $\hat{\vec{e}}^\pm_i =  \hat{\vec{e}}^x_i \pm \mathrm{i} \hat{\vec{e}}^y_i$, and $\{\hat{\vec{e}}^x_i,\hat{\vec{e}}^y_i, \hat{\vec{e}}^z_i\}$ the local orthogonal basis, with $\hat{\vec{e}}^z_i$ along the ground state direction. The linear-in-magnons part of the magnetic moment thus reads
\begin{align}
    \vec{m}_i^{(1)} = \vec{\mu}_i b_i + \text{H.c.},
    \qquad 
    \vec{\mu}_i = -\mathsf{g} \mu_\text{B} \sqrt{S/2} \hat{\vec{e}}^-_i,
    \label{eq:m1_expanded}
\end{align}
where H.c.~denotes the Hermitian conjugate.
We assume a spin-driven electric polarization, such that $\vec{p}_i$ can be expanded in terms of spin operators \cite{Moriya1968} and, hence, in magnon operators,
\begin{align}
    \vec{p}_i = \vec{p}_i^{(0)} + \vec{p}_i^{(1)}  + \vec{p}_i^{(2)} + \ldots ,
    \label{eq:expansion-of-p}
\end{align}
where $\vec{p}_i^{(l)}$ contains $l$ magnon operators. While $\vec{p}_i^{(0)}$ is the ground state electric moment, the linear-in-magnons contribution,
\begin{align}
    \vec{p}_i^{(1)} = \vec{\pi}_i b_i + \text{H.c.},
    \label{eq:p1_expanded}
\end{align}
encompasses dynamical fluctuations and $\vec{\pi}_i$ is a material-dependent expansion coefficient. It can be nonzero and allow electrically active excitations even if $\vec{p}_i^{(0)} = 0$ \cite{tokuraMultiferroicsSpinOrigin2014a}. The bilinear $\vec{p}_i^{(2)}$ encodes the electric dipole moment expectation value of the magnetic fluctuations.

For the electromagnetic fields, we perform a standard photon quantization, such that the field operators in the Schr\"{o}dinger picture read
\begin{subequations}
\begin{align}
	\vec{E}(\vec{r}_i)
	&=
	\sum_{\beta} \left( 
	\vec{\mathcal{E}}_{\beta}(\vec{r}_i) a_\beta
	+ \vec{\mathcal{E}}_{\beta}^*(\vec{r}_i) a^\dagger_\beta
	\right),
	\label{eq:e_field_quantized}
	\\
	\vec{B}(\vec{r}_i)
	&=
	\sum_{\beta} \left( 
	\vec{\mathcal{B}}_{\beta}(\vec{r}_i) a_\beta
	+ \vec{\mathcal{B}}_{\beta}^*(\vec{r}_i) a^\dagger_\beta
	\right),
	\label{eq:b_field_quantized}
\end{align}
\end{subequations}
where $\beta$ is the index of the mode created (annihilated) by the bosonic operator $a^\dagger_\beta$ ($a_\beta$) and $\vec{\mathcal{E}}_{\beta}(\vec{r}_i)$ and $\vec{\mathcal{B}}_{\beta}(\vec{r}_i)$ are the associated complex electric-field and magnetic-field vacuum fluctuations, respectively.
We restrict our attention to a single microwave mode of energy $\hbar \Omega$, where $\hbar$ is the reduced Planck's constant. Additionally, we assume that the linear scale of the sample is much smaller than the photon wavelength, such that the fields in the sample can be treated as spatially uniform. Below, we drop the $\beta$ index and set $\vec{\mathcal{E}}_{\beta}(\vec{r}_i) = \vec{\mathcal{E}}$ and $\vec{\mathcal{B}}_{\beta}(\vec{r}_i) = \vec{\mathcal{B}}$, such that the cavity Hamiltonian effectively contains a single mode, $\mathcal{H}_\text{CA}=\hbar\Omega a^\dagger a$. 
Thus, to lowest order in operators, the dynamical coupling between the electromagnetic fields and the material reads
\begin{align}
	\mathcal{H}^{(1)}_\text{EM}
	= 
	- \sum_{i=1}^{N} 
	[
	( a \vec{\mathcal{E}} + a^\dagger \vec{\mathcal{E}}^* ) \cdot \vec{\pi}_i
	+
	( a \vec{\mathcal{B}} + a^\dagger \vec{\mathcal{B}}^* ) \cdot \vec{\mu}_i ] b_i
	+
	\text{H.c.}
\end{align}

We now assume that there are in total $N_\text{u}$ magnetic unit cells, each of which features $N_\text{b}$ basis spins, such that $N_\text{u} \times N_\text{b} = N$. The Fourier transform of the magnon operators reads
\begin{align}
	b_{i,j}^\dagger = \frac{1}{\sqrt{N}_\text{u}} \sum_{\vec{k}} \mathrm{e}^{-\mathrm{i} \vec{k} \cdot \vec{r}_{i,j}} b^\dagger_{\vec{k},j},
\end{align} 
where $i$ labels the unit cell and $j$ the basis atom.
We perform a Bogoliubov transformation to magnon normal modes, i.e.,
\begin{align}
    \begin{pmatrix}
        b_{\vec{k},j} \\ b^\dagger_{-\vec{k},j}
    \end{pmatrix}
    =
    \sum_{n=1}^{N_\text{b}}
    T^{j,n}_{\vec{k}}
    \begin{pmatrix}
        b_{\vec{k},n} \\ b^\dagger_{-\vec{k},n}
    \end{pmatrix},
    \quad
    T^{j,n}_{\vec{k}}
    =
    \begin{pmatrix}
        u^{j,n}_{\vec{k}} & \left(v^{j,n}_{-\vec{k}} \right)^*\\
        v^{j,n}_{\vec{k}} & \left(u^{j,n}_{-\vec{k}}\right)^*
    \end{pmatrix},
    \label{eq:Bogo}
\end{align}
where $n$ is the band index and $T^{j,n}_{\vec{k}}$ part of the paraunitary matrix that diagonalizes the bilinear magnon Hamiltonian \cite{Colpa1978}.
Only the $\vec{k} = 0$ magnon gives rise to finite coupling,
\begin{align}
	\mathcal{H}^{(1)}_\text{EM}
	&= 
	\sum_{ n=1 }^{N_\text{b}} 
	\left(
	    \hbar g^{(n)}
	    a^\dagger b_{0,n}
    	+
    	\hbar \tilde{g}^{(n)}
	    a b_{0,n}
    	+
	    \text{H.c.}
	\right),\label{eq:ME coupling}
\end{align}
with mode-dependent MP coupling constants
\begin{subequations}
\begin{align}
    \hbar g^{(n)} &= \hbar g^{(n)}_\text{m-el} + \hbar g^{(n)}_\text{m-mag},
    \\
    \hbar \tilde{g}^{(n)} &= \hbar \tilde{g}^{(n)}_\text{m-el} + \hbar \tilde{g}^{(n)}_\text{m-mag}.
\end{align}
\end{subequations}
Their electric and magnetic components are given by
\begin{subequations}
\begin{align}
    \hbar g^{(n)}_\text{m-el}
    &=
    - \sqrt{N} 
        \vec{\mathcal{E}}^* \cdot \vec{\CalP}^{(n)}
    ,
    \;
    \hbar g^{(n)}_\text{m-mag}
    =
    - \sqrt{N} 
        \vec{\mathcal{B}}^* \cdot \vec{\CalM}^{(n)},
	\label{eq:EMcoup}
	\\
	\hbar \tilde{g}^{(n)}_\text{m-el}
    &=
    - \sqrt{N} 
        \vec{\mathcal{E}} \cdot \vec{\CalP}^{(n)}
    ,
    \;
    \hbar \tilde{g}^{(n)}_\text{m-mag}
    =
    - \sqrt{N} 
        \vec{\mathcal{B}} \cdot \vec{\CalM}^{(n)}.
\end{align}
\end{subequations}
We defined the dynamic magnetic and electric dipole moments of the $n$th magnon mode as
\begin{subequations}
\begin{align}
    \boldsymbol{\CalP}^{(n)}&= \frac{1}{\sqrt{N_\text{b}}} \sum_{ j=1 }^{N_\text{b}} \left( \vec{\pi}_{j} u^{j,n}_{0} + \vec{\pi}^*_{j} v^{j,n}_{0} \right),
    \label{eq:MagnonP}\\
    \boldsymbol{\CalM}^{(n)}&= \frac{1}{\sqrt{N_\text{b}}} \sum_{ j=1 }^{N_\text{b}} \left(\vec{\mu}_{j} u^{j,n}_{0} + \vec{\mu}^*_{j} v^{j,n}_{0} \right),
    \label{eq:MagnonM}
\end{align}
\end{subequations}
where $u_{\vec{k}}^{j,n}$~($v_{\vec{k}}^{j,n}$) is the particle (hole) sector of the wave function of the $n$th magnon mode [cf.~Eq.~\eqref{eq:Bogo}]. We provide additional information on the dynamic moments in Appendix~\ref{sec:multipole_magnon}.

\subsection{Symmetries of magnetoelectric magnon-photon coupling}
Magnetoelectricity is tied to a breaking of time-reversal symmetry $\mathcal{T}$ and space inversion symmetry $\mathcal{I}$. Hence, application of either of the two operations is associated with a change of the MP coupling $\hbar g^{(n)}$ (and $\hbar \tilde{g}^{(n)}$). Space inversion $\mathcal{I}$ leaves the magnetic domain unaffected, $\mathcal{I} \boldsymbol{\CalM}^{(n)} \to \boldsymbol{\CalM}^{(n)}$, but flips the microscopic polarization vectors, resulting in $\mathcal{I} \boldsymbol{\CalP}^{(n)} \to - \boldsymbol{\CalP}^{(n)}$. In contrast, time-reversal $\mathcal{T}$ comes with a complex conjugation due to its antiunitarity and a reversal of the magnetic moments, $ \mathcal{T} \boldsymbol{\CalM}^{(n)} \to - (\boldsymbol{\CalM}^{(n)})^*$ and $\mathcal{T} \boldsymbol{\CalP}^{(n)} \to (\boldsymbol{\CalP}^{(n)})^*$. Both operations flip the sign of exactly one of the two moments, such that, in general, 
\begin{subequations}
\begin{align}
    |\hbar g^{(n)}(\vec{M},\vec{P})| &\ne |\hbar g^{(n)}(-\vec{M},\vec{P})|,
    \\
   |\hbar g^{(n)}(\vec{M},\vec{P})| &\ne |\hbar g^{(n)}(\vec{M},-\vec{P})|.
\end{align}
\end{subequations}
Only the combined operation is a symmetry,
\begin{align}
   |\hbar g^{(n)}(\vec{M},\vec{P})| = |\hbar g^{(n)}(-\vec{M},-\vec{P})|.
\end{align}

\section{Worked Example: Multiferroic insulator  C\titlelowercase{u}$_2$OS\titlelowercase{e}O$_3$}
\label{sec: worked_example}
To explore the experimental implications of the ME coupling $\hbar g^{(n)}$, we consider the multiferroic material Cu$_2$OSeO$_3$ as an example~\cite{sekiObservationSkyrmionsMultiferroic2012, White2012, White2014}.
As shown in Ref.~\cite{sekiMagnetoelectricNatureSkyrmions2012}, in this material, the spin-driven electric polarization is brought about by the microscopic p-d hybridization mechanism \cite{Jia2006, Jia2007, arimaFerroelectricityInducedProperScrew2007}.
We work in the approximation of effective $S=1$ spins per Cu tetrahedron~\cite{mochizukiDynamicalMagnetoelectricPhenomena2015}. As a result, the (coarse-grained) local electric moments per effective tetrahedron spin can be written as~\cite{mochizukiDynamicalMagnetoelectricPhenomena2015, arimaFerroelectricityInducedProperScrew2007, sekiMagnetoelectricNatureSkyrmions2012,  belesiMagnetoelectricEffectsSingle2012}
\begin{align}
    \vec{p}_i = \frac{\lambda}{2} 
        \begin{pmatrix}
            S^b_i S^c_i + S^c_i S^b_i \\   
            S^c_i S^a_i + S^a_i S^c_i \\
            S^a_i S^b_i + S^b_i S^a_i 
        \end{pmatrix}
        \label{eq:pCuOSeO}
\end{align}
in symmetrized form, with $\lambda = \unit[5.64 \times 10^{-27}]{\mu C \, m}$, obtained at $5\,$K under the magnetic field parallel to [110] direction \cite{mochizukiDynamicalMagnetoelectricPhenomena2015}. $\lambda$ is a pseudoscalar that changes sign under space inversion. 
The $a,b,c$ superscripts denote the crystallographic axes. 
Let us assume that the $i$th spin in the classical ground state is parametrized as
\begin{align}
    \langle \vec{S}_i \rangle 
    = 
    S
    (\cos \phi_i \sin \theta_i, \sin \phi_i \sin\theta_i,\cos\theta_i)^\text{T}   
\end{align}
by angles $\phi_i$ and $\theta_i$. 
A magnon expansion yields 
$
    \vec{p}_i \approx \vec{p}_i^{(0)} + (\vec{\pi}_i b_i + \text{H.c.})
$,
where the static and dynamical moments respectively read
\begin{subequations}
\begin{align}
    \vec{p}_i^{(0)} &= \lambda S^2
        \begin{pmatrix}
             \cos \theta_i \sin \theta_i \sin \phi_i \\
              \cos\theta_i \cos \phi_i \sin \theta_i \\
              \cos \phi_i \sin^2\theta_i \sin\phi_i
        \end{pmatrix},
        \label{eq:GS-polarization}
    \\
	\vec{\pi}_i 
	&= 
	\frac{\lambda S^{\frac{3}{2}}}{\sqrt{2}}
	\begin{pmatrix}
		\cos (2\theta_i) \sin \phi_i
		-\mathrm{i} \cos \theta_i \cos \phi_i 
		\\
		\cos (2\theta_i) \cos \phi_i
		+\mathrm{i}\cos \theta_i \sin \phi_i
		\\
		\sin \theta_i \cos \theta_i \sin (2\phi_i)
		- \mathrm{i} \sin \theta_i  \cos ( 2 \phi_i )
	\end{pmatrix}.
	\label{eq: dynamical_pi}
\end{align}
\end{subequations}

\begin{figure}[t]
    \centering
    \includegraphics[width=1\columnwidth]{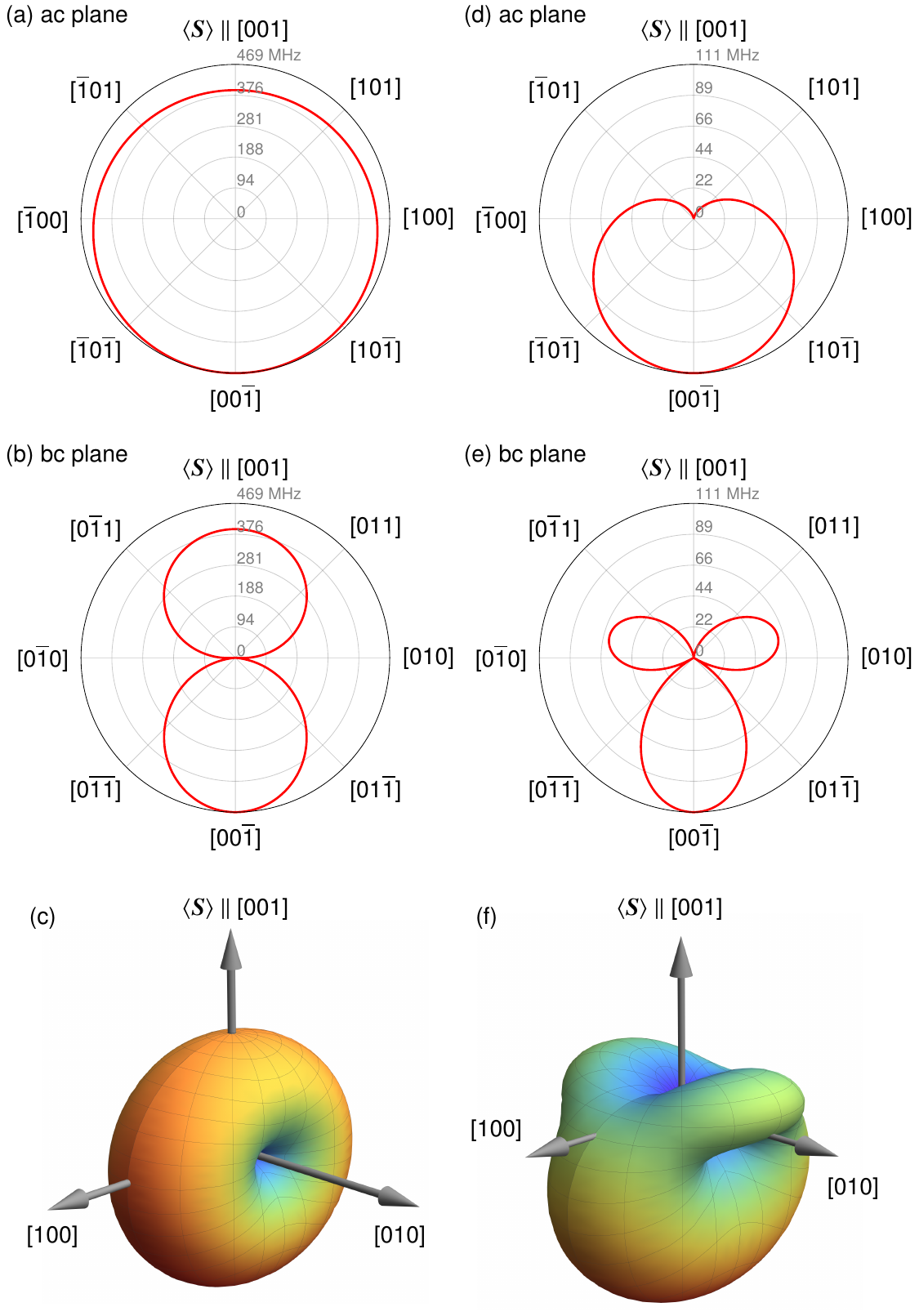}
    \caption{ 
        Evolution of the magnetoelectric magnon-photon coupling $|g|/2\pi$ in the ferromagnetic phase of Cu$_2$OSeO$_3$ upon rotation of the spin moment $\langle \vec{S} \rangle$ in (a,d) the ac plane, (b,e) the bc plane, and (c,f) a general direction. Notice in particular the asymmetry with respect to magnetization reversal, $|\hbar g(\langle \vec{S} \rangle )| \ne |\hbar g(-\langle \vec{S} \rangle )|$.
        Cavity field directions read $\vec{\mathcal{E}} \parallel [100]$ and $\vec{\mathcal{B}} \parallel [010]$. Field amplitudes are chosen as (a-c) $\mathcal{E} = \mathcal{E}_0/\sqrt{2}$ and $\mathcal{B}=\mathcal{B}_0/\sqrt{2}$ and (d-f) $\mathcal{E} = 0.996 \mathcal{E}_0$ and $\mathcal{B} = 0.091 \mathcal{B}_0$.
    }
    \label{fig:angulardependence}
\end{figure}

\subsection{Field-polarized phase}
We first consider the field-polarized phase of Cu$_2$OSeO$_3$ and set $\langle \vec{S}_i \rangle = \langle \vec{S} \rangle \parallel [001]$. We drop the band index $n$ ($N_\text{b} = 1$), such that the effective spin-lattice Hamiltonian reads $\mathcal{H}_\text{SL}= \hbar \omega_\text{m} b^\dagger b$ within linear spin-wave theory, where $\hbar \omega_\text{m}$ is the magnon energy. Although $\vec{p}_i^{(0)} = 0$, the dynamical polarization is finite, 
$
    \vec{\pi}_i = \vec{\pi} = 
    \lambda S^{3/2}/ \sqrt{2} ( - \mathrm{i},1,0)^\text{T},
$
and allows a coupling with electric fields in the $ab$ plane; e.g.
$
    | \hbar g_\text{m-el} | = \sqrt{N/2} \lambda S^{3/2} \mathcal{E}
$
for $\vec{\mathcal{E}} \parallel [100]$.
With $\mathcal{E} = \mathcal{E}_0 \equiv c \sqrt{\hbar \Omega \mu_0/(2V)}$ follows 
$
    | \hbar g_\text{m-el} | 
    = 
    \lambda c \sqrt{\hbar \Omega \mu_0 \rho_\text{s} \eta}  / 2
$,
where $\Omega$ is the photon frequency, $c$ the speed of light, $\mu_0$ the vacuum permeability, $\rho_\text{s} = \unit[5.67 \times 10^{27}]{m^{-3}}$ the spin density of Cu$_2$OSeO$_3$~\cite{mochizukiDynamicalMagnetoelectricPhenomena2015}, and $\eta = V_\text{crys} / V$ the ratio of crystal and cavity volume.
Taking $\Omega/(2\pi) = \unit[9.74]{GHz}$ and $\eta = 0.041$ \cite{Abdurakhimov2019}, we obtain
\begin{align}
    \frac{| g_\text{m-el} |}{2\pi}
    =
    \unit[55.4]{MHz},
    \label{eq:electrical-g}
\end{align}
which is larger than the typical magnetic damping $\gamma_\text{m}/(2\pi) \approx \unit[50]{MHz}$ \cite{Abdurakhimov2019} (full width at a half maximum), suggesting that it is possible to reach the strong coupling limit purely electrically. With a photonic damping rate of $\kappa/(2\pi) = \unit[1]{MHz}$, we find the cooperativity $C = | g_\text{m-el} |^2 / (\gamma_\text{m} \kappa) \approx 61.4$. 

Let us compare the electric coupling to the usual magnetic dipole coupling
$
    | \hbar g_\text{m-mag} | 
    = 
    \mathsf{g} \mu_\text{B} \mathcal{B} \sqrt{N/2}
$
for $\vec{\mathcal{B}} \parallel [010]$ and
$
    \mathcal{B} = \mathcal{B}_0 \equiv \sqrt{\hbar \Omega \mu_0/(2V)}
$, 
yielding 
\begin{align}
    \frac{|g_\text{m-mag}|}{2\pi} = \unit[607.9]{MHz}.
\end{align}
Thus, for Cu$_2$OSeO$_3$, the maximal magnetic coupling is stronger by a factor of
\begin{align}
    \frac{| g_\text{m-mag}|}{|g_\text{m-el} |}
    = 
    \frac{\mathsf{g} \mu_\text{B}}{\lambda} \cdot \frac{\mathcal{B}_0}{\mathcal{E}_0}
    \approx 
    11,
    \label{eq:ratio}
\end{align}
which is not surprising given that the electric polarization derives from the $p$-$d$ hybridization mechanism which is a second-order relativistic effect~\cite{arimaFerroelectricityInducedProperScrew2007,arimaSpinDrivenFerroelectricityMagnetoElectric2011}.

We now study the interplay of magnetic and electric couplings. We fix $\vec{\mathcal{E}} \parallel [100]$ and $\vec{\mathcal{B}} \parallel [010]$ and focus on the magnetization direction dependence of $|\hbar g|$ (see Fig.~\ref{fig:angulardependence}; recall $\langle \vec{S} \rangle \parallel - \vec{M}$).
As advertised above, a reversal of the magnetization direction leads to a change of the MP coupling, explicitly,
$
    |\hbar g|
    =\sqrt{N} |\mathrm{i} \sqrt{2} \mathsf{g} \mu_\text{B} \mathcal{B} \mp \mathrm{i} \lambda \mathcal{E} |/2
$,
where $-$ ($+$) applies for $\langle \vec{S} \rangle \parallel [001]$ ($\langle \vec{S} \rangle \parallel [00\overline{1}]$). For $\mathcal{E} = \mathcal{E}_0/\sqrt{2}$ and $\mathcal{B} = \mathcal{B}_0/\sqrt{2}$, rotating $\langle \vec{S} \rangle$ from the $[001]$ to the $[00\overline{1}]$ direction amounts to a change from $|g|/(2\pi) = \unit[391]{MHz}$ to $|g|/(2\pi) =\unit[469]{MHz}$, which is an increase by about $20\,\%$ [cf.~Figs.~\ref{fig:angulardependence}(a-c)]. 
If the sample is placed at a point in the cavity where the ratio between $\mathcal{E}$ and $\mathcal{B}$ compensates for the factor in Eq.~\eqref{eq:ratio}, that is, $\mathcal{E} \approx 0.996 \mathcal{E}_0$ and $\mathcal{B} \approx 0.091 \mathcal{B}_0$, electric and magnetic couplings can become identical in strength. The resulting maximally anisotropic coupling shown in Figs.~\ref{fig:angulardependence}(d-f) allows, in particular, a vanishing coupling for $\langle \vec{S} \rangle \parallel [001]$.

We propose two strategies for the experimental detection of the ME MP coupling by measuring the associate spectral splitting in the ferromagnetic phase of Cu$_2$OSeO$_3$: 
(1) The pure electric coupling can be probed by either placing the sample at a node of the magnetic field or by aligning the magnetic field with the magnetization direction. Any observed MP splitting must be attributed to the coupling to the polarization.
(2) The predicted asymmetry $|\hbar g(\vec{M},\vec{P})| \ne |\hbar g(-\vec{M},\vec{P})|$ may be probed straight-forwardly by placing the magnet in finite electric and magnetic fields and comparing splittings for opposite magnetization directions.

\subsection{Skyrmion crystal phase}
\label{sec: SkX}

Strictly speaking, the period of skyrmion spin textures is incommensurate with the period of crystal structures. However, the size of skyrmions in Cu$_2$OSeO$_3$ is approximately 70 times larger than the lattice constant~\cite{sekiObservationSkyrmionsMultiferroic2012}, justifying the use of a
continuum approximation. Here, we consider a two-dimensional spin lattice Hamiltonian discretized on the triangular lattice as a minimal model~\cite{mochizukiSpinWaveModesTheir2012}: 
\begin{eqnarray}
\label{eq: SpinLatticeH}
\mathcal{H}_\text{SL}&=&\frac{1}{2}\sum_{\braket{\vec{r},\vec{r}'}}(-J_{\vec{r},\vec{r}'}\vec{S}_{\vec{r}}\cdot \vec{S}_{\vec{r}'}+ \vec{D}_{\vec{r},\vec{r}'}\cdot\vec{S}_{\vec{r}}\times\vec{S}_{\vec{r}'})\nonumber\\
&+&\mathsf{g}\mu_{B} B_z \sum_{\vec{r}} \vec{S}_{\vec{r}}\cdot\hat{\vec{z}} \,,
\end{eqnarray}
where $\braket{\vec{r},\vec{r}'}$ is summed over nearest neighbors of a triangular lattice with the ferromagnetic exchange interaction $J>0$ and Dzyaloshinskii-Moriya interaction $\vec{D}_{{\vec{r},\vec{r}'}}=D(\vec{r}-\vec{r}')/|\vec{r}-\vec{r}'|$. The unit vector $\hat{\vec{z}}$ defines the out-of-plane direction along the external magnetic field $B_z$. In Cu$_2$OSeO$_3$, SkXs were stabilized for $\hat{\vec{z}}\parallel $[001], [110], and [111]~\cite{sekiObservationSkyrmionsMultiferroic2012}. 
Although we take the ratio $D/J=0.5$ to reduce computational costs, the energies and magnetic fields are rescaled for the more realistic parameter $D/J=0.09$ in Appendix~\ref{sec: rescaling}. 
The SkX ground state, as shown in Fig.~\ref{fig:teaser}, is obtained at a finite magnetic field by combining Monte Carlo and Landau-Lifshitz-Gilbert annealing~\cite{evansAtomisticSpinModel2014}.
The magnon band spectrum is obtained within linear spin-wave theory~\cite{RoldanMolina2016,Diaz2019AFM, Diaz2020FM,Hirosawa2020,Mook2020QuantumDamping}, as detailed in Appendix~\ref{sec: spinwave_skx}.

Magnon modes of SkXs can be characterized by dynamic magnetic dipole moments. 
For example, the counterclockwise (CCW) and clockwise (CW) modes possess $\CalM_x$ and $\CalM_y$, while the breathing mode possesses $\CalM_z$, which was confirmed by the selection rule for the microwave spectroscopy~\cite{mochizukiSpinWaveModesTheir2012, onoseObservationMagneticExcitations2012, schwarzeUniversalHelimagnonSkyrmion2015,Garst2017}.
In addition to these magnetically active modes, SkXs support polygon deformation modes, arising from single-skyrmion bound states~\cite{linInternalModesSkyrmion2014, Schutte2014, Diaz2020FM, diazSpinWaveRadiation2020a}. These modes are characterized by azimuthal quantum number $\ell$~\cite{shekaInternalModesMagnon2001}, with $\ell=2$ for elliptic, $\ell=3$ for triangle, $\ell=4$ for square, $\ell=5$ for pentagon, and $\ell=6$ for hexagon, showing higher multipolar characters. Crucially, Eq.~\eqref{eq:pCuOSeO} relates magnetic quadrupole moments with electric dipole moments in Cu$_2$OSeO$_3$. As a result, the elliptic mode, which carries magnetic quadrupole moment $Q_{xy}$~\cite{Schutte2014}, exhibits a non-zero electric polarization $\CalP_z$ for $\hat{\vec{z}}\parallel [001]$, rendering it electrically active.

\begin{figure}[t]
    \centering
    \includegraphics[width=\columnwidth]{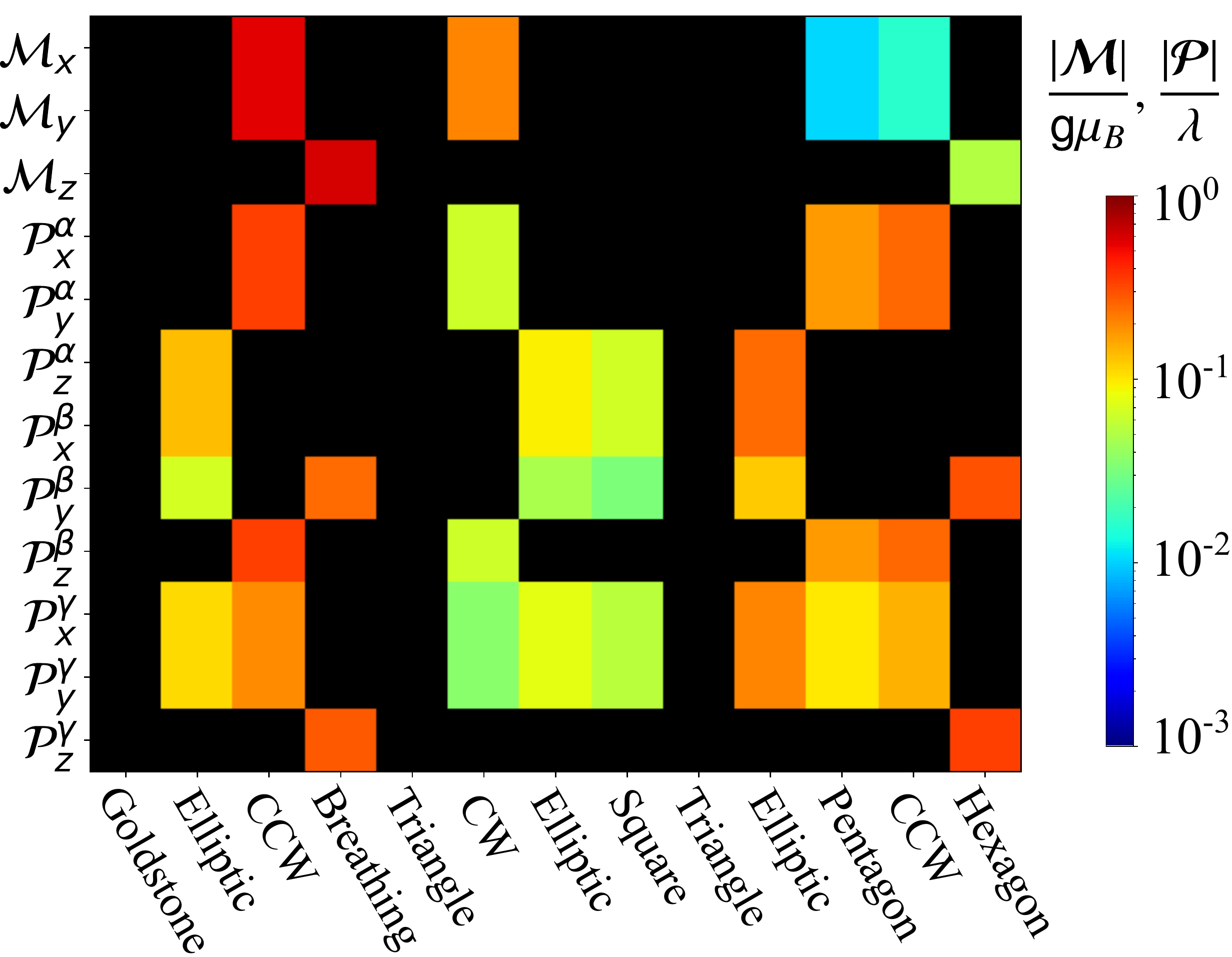}
    \caption{Dynamical magnetic and electric dipole moments of low-energy magnon modes of skyrmion crystals in Cu$_2$OSeO$_3$. The dimensionless amplitudes of $\vec{\CalM}/(\mathsf{g}\mu_B)$ and $\vec{\CalP}/\lambda$---plotted on a logarithmic scale---were obtained at $D/J=0.5$ and $\mathsf{g} \mu_B B_z J / (D^2S)=0.8$. We denote $\boldsymbol{\CalP}^{\alpha},\, \boldsymbol{\CalP}^{\beta}, \textrm{ and }\boldsymbol{\CalP}^{\gamma}$ as electric polarization with $\hat{\vec{z}}\parallel [001]$, $[110]$, and $[111]$, respectively~(cf.~Appendices~\ref{sec:dynamical_SkX} and~\ref{sec:multipole_SkX}). For all cases, the $x$-axis is taken along $[\overline{1}10]$. Black regions indicate dipole moments smaller than $10^{-3}$. 
    }
    \label{fig: dipole}
\end{figure}

Using Eqs.~\eqref{eq:MagnonP} and~\eqref{eq:MagnonM}, we compute $\vec{\CalM}$ and $\vec{\CalP}$ of low-energy magnon modes in SkXs for three experimentally relevant directions of the static magnetic field as shown in Fig.~\ref{fig: dipole}.
(For additional information on the necessary rotation of the coordinate system and magnetic quadrupoles, see Appendices~\ref{sec:dynamical_SkX} and~\ref{sec:multipole_SkX}, respectively.)
Besides the three magnetically active CCW, CW, and breathing modes, we obtain large $\vec{\CalP}$ in the elliptical modes due to their magnetic quadrupole moment.  Furthermore, the other polygon modes carry nonzero $\vec{\CalM}$ and $\vec{\CalP}$, which is not allowed if $\ell$ is a good quantum number, as is the case for a single skyrmion. However, for SkXs, we find that the skyrmion-skyrmion interaction with $C_6$ rotational symmetry breaks the rotational symmetry of polygon modes, resulting in strong hybridization with magnetically and/or electrically active modes
, except for the triangle (sextupole) modes that are ``magnetoelectrically dark.'' 
(For a detailed numerical investigation, see Appendix~\ref{sec:hybridizatin}.)
Instead, it takes a small cubic anisotropy---which was neglected here---for the triangle modes to hybridize with other modes~\cite{aqeelMicrowaveSpectroscopyLowTemperature2021b, takagiHybridizedMagnonModes2021}.

\begin{figure}[t]
    \centering
    \includegraphics[width=\columnwidth]{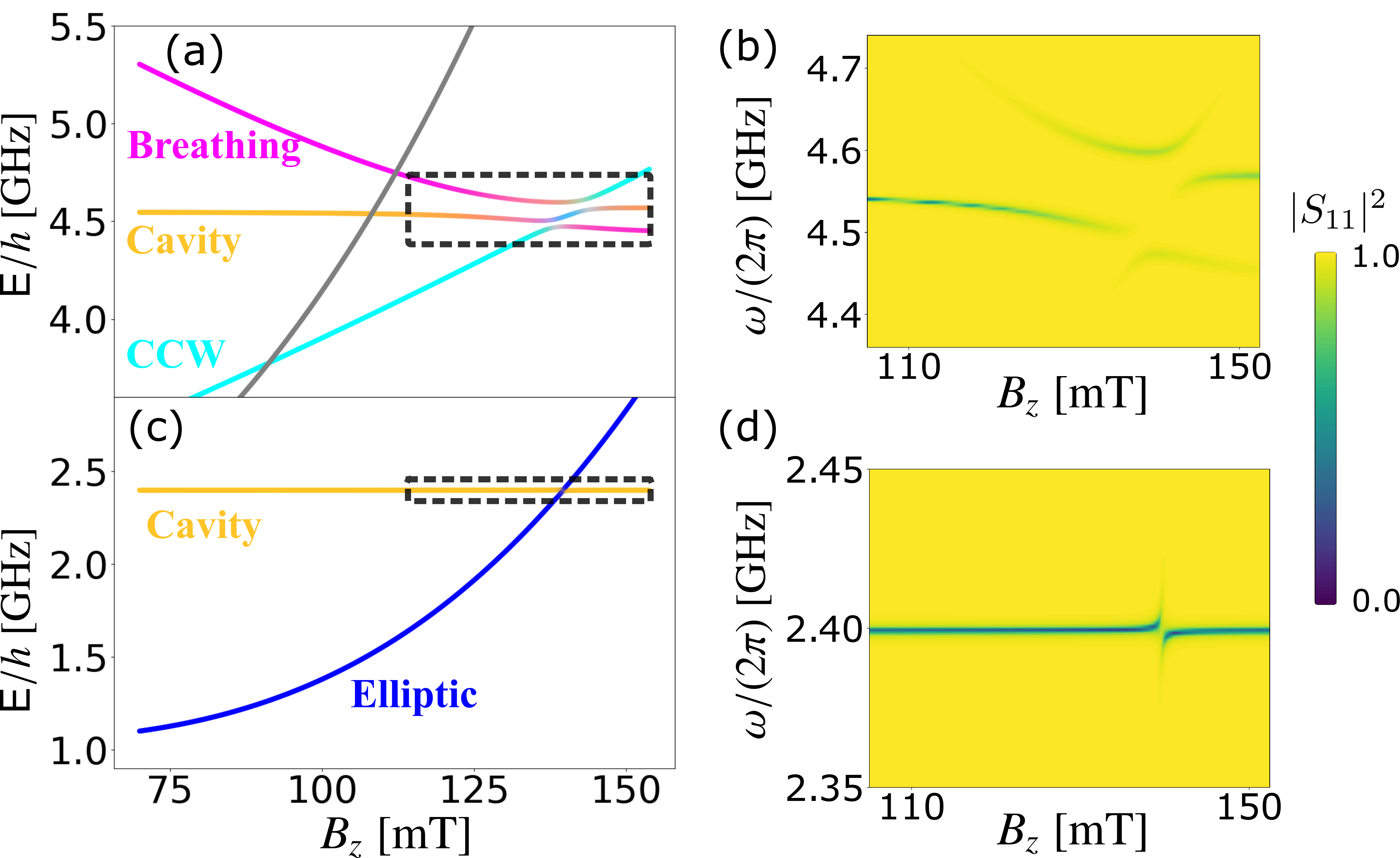}
    \caption{Cavity-induced magnon-magnon interaction and MP hybridization of the magnetically dark elliptic mode. 
    (a,c) Magnetic field dependence of magnon-photon spectrum of the low temperature SkX, obtained at $D/J=0.5$ and $\hat{\vec{z}}\parallel [001]$. The Planck constant is denoted as $h$. The color code indicates the probability density of cavity mode~(dark yellow) and skyrmion excitations including breathing~(magenta), CCW~(cyan), and elliptic~(blue) modes. (b,d)~Calculated microwave reflection $|S_{11}(\omega)|^2$ using the input-output formalism~\cite{Abdurakhimov2019,walls2008quantum} (cf.~Appendix~\ref{sec:experiment_SkX}), obtained for regions enclosed by black dashed lines in (a) and (c). 
    The cavity field directions are fixed as (a,b)~$\vec{\CalE}\parallel [\overline{1}10]$ and $\vec{\CalB}\parallel [001]$ at $\Omega/(2\pi)=4.55$~GHz and (c,d)~$\vec{\CalE}\parallel [001]$ and $\vec{\CalB}\parallel [\overline{1}10]$ at $\Omega/(2\pi)=2.4$~GHz. In (a-d), we assume that the SkX is placed slightly away from the antinode of the electric component of the cavity mode with $|\vec{\CalB}|\approx 0.1\CalB_0$ and $|\vec{\CalE}|\approx 0.995\CalE_0$.
    We also assume the damping rate of cavity mode $\kappa_a/(2\pi)=1$~MHz, the damping rate of the $n$th magnon mode $\kappa_n/(2\pi)=20$~MHz, and $\eta=0.1$. The energy eignenvalues $\mathsf{E}$ and magnetic fields are rescaled for Cu$_2$OSeO$_3$ at $5$~K~(cf.~Appendix~\ref{sec: rescaling}). 
    }
    \label{fig: S11_plot}
\end{figure}

For experimental studies of ME MP coupling, the low-temperature SkX phase provides an ideal platform, which can be realized in a thin film or quenched bulk crystal~\cite{sekiObservationSkyrmionsMultiferroic2012,aqeelMicrowaveSpectroscopyLowTemperature2021b,takagiHybridizedMagnonModes2021}. Noting a remarkably small Gilbert damping and enhanced effective exchange interaction at low temperatures~\cite{mochizukiDynamicalMagnetoelectricPhenomena2015,stasinopoulosLowSpinWave2017}, the strong coupling limit can be realized with the magnetic coupling constant for the CCW and breathing modes potentially exceeding $500$~MHz~(cf.~Appendix~\ref{section:coupling_strength} for the the numerical evaluation of MP coupling in skyrmion eigenmodes). Furthermore, the ME coupling leads to fundamentally new phenomena when the damping rate of skyrmion eigenmodes becomes smaller than $100$~MHz. 
Let us consider the SkX placed slightly away from the antinode of the electric component of the cavity
mode with $\hat{\vec{z}}\parallel [001]$, where $|\vec{\CalB}|\approx 0.1\CalB_0$ and $|\vec{\CalE}|\approx 0.995\CalE_0$. Applying $\vec{\CalE}\parallel [\overline{1}10]$ and $\vec{\CalB}\parallel [001]$ at $\Omega/(2\pi)=4.55$~GHz, the triple resonance condition is satisfied with the cavity mode coupled magnetically and electrically with the breathing and CCW modes, respectively. As demonstrated in Figs.~\ref{fig: S11_plot}(a,b), the cavity-mediated magnon-magnon coupling induces a hybridization gap between the breathing and CCW modes, which can be experimentally observed in microwave reflection $|S_{11}(\omega)^2|$.
(For the theory of microwave reflection and additional discussion of MP hybridizations, see  Appendix~\ref{sec:experiment_SkX}.)
We also find a small anticrossing between the cavity mode and elliptic mode of purely electrical origin, as shown in Fig.~\ref{fig: S11_plot}(c) and~(d), hence allowing the MP hybridization of magnetically dark modes.

\section{Quantum information applications of magnetoelectric magnon-photon coupling}
\label{sec: QI_application}

\subsection{Electrically tunable magnon-mediated coherent information transduction between microwave photons}
\label{sec: QI_application1}
As reviewed in Ref.~\cite{Li2020HybridMagnonics}, magnon-photon coupling may find application in scalable hybrid quantum computing architectures, e.g., in quantum information processing with on-chip magnonics and superconducting circuits where it is used to implement quantum gates~(see Appendix~\ref{sec:unitary_gate} for magnon-mediated and photon-mediated quantum gates). One may use a magnet to couple two circuit quantum electrodynamics systems, each of which consists of a superconducting qubit and a microwave cavity. As a result, quantum information can be controllably transduced between the two systems because they effectively interact with each other via the tunable magnon-photon coupling. The tunability of the latter arises from controlling the magnon frequency by external means.

In contrast to the well-established magnetic-field tunability of the magnon frequency, $\omega_\text{m} = \omega_\text{m}(B)$, magnetoelectricity facilitates an additional electric-field tunability, $\omega_\text{m} = \omega_\text{m}(E)$.  The electric-field dependence of the magnon frequency can be extracted from the bilinear part $\vec{p}_i^{(2)}$ in the magnon expansion Eq.~\eqref{eq:expansion-of-p} of the local electric dipole moments. As shown in Appendix \ref{sec:p2}, in the collinear ferromagnetic phase of Cu$_2$OSeO$_3$, the electric-field induced magnon frequency shift can be written as 
$
    \Delta f = \unit[14.7]{Hz} \times \frac{E}{\unit{V/m}}
$,
where we assumed $\langle \vec{S} \rangle, \vec{E} \parallel [111]$. Thus, similar to existing protocols of magnetic-field pulses \cite{trevillian2020unitary,*trevillian2021unitary, Li2020HybridMagnonics, Awschalom2021}, it is feasible to design electric-field pulse protocols to tune the magnon frequency relative to photon frequencies and to realize quantum operations like SPLIT and SWAP operations between two different photon modes. 


Below, we explore the possibility of realizing SPLIT and SWAP operations by all-electrical means to highlight the importance of magnetoelectric coupling. We emphasize that ``all-electrical'' refers to both the magnon frequency tuning and the magnon-photon coupling being a result of electric dipole coupling. We model one magnon mode and two photon modes with the respective complex amplitudes given by $b(t)$, $a_1(t)$, and $a_2(t)$. Their time evolution is governed by the equation of motion \cite{trevillian2020unitary,*trevillian2021unitary, Li2020HybridMagnonics, Awschalom2021}
\begin{align}
    \begin{pmatrix}
      \dot{a}_1(t) \\ \dot{a}_2(t) \\ \dot{b}(t)
    \end{pmatrix}
    =
    \begin{pmatrix}
        -\mathrm{i} \Omega_1 - \frac{\kappa_1}{2} & 0 & -\mathrm{i} g_1 \\
        0 & -\mathrm{i} \Omega_2 - \frac{\kappa_2}{2} & -\mathrm{i} g_2 \\
        -\mathrm{i} g_1 & - \mathrm{i} g_2 &  -\mathrm{i} \omega_\text{m}(t) - \frac{\gamma_\text{m}}{2} 
    \end{pmatrix}
    \begin{pmatrix}
      a_1(t) \\ a_2(t) \\ b(t)
    \end{pmatrix},
    \label{eq:EOM}
\end{align}
where $\Omega_1$ and $\Omega_2$ are the photon frequencies, $\omega_\text{m}(t) = \omega_\text{m}[E(t)]$ is the $E$-field dependent magnon frequency, $\kappa_1$ and $\kappa_2$ are the photon damping, $\gamma_\text{m}$ is the magnon damping (full width at a half maximum), and $g_1$ and $g_2$ are the magnon-photon coupling of the first and second photon, respectively. The potential coupling between cavity modes via an output transmission line is neglected.
We set $\kappa_1 = \kappa_2 = 0$ for simplicity, assume a high-quality crystal with low magnon damping $\gamma_\text{m}/(2\pi) = \unit[8]{MHz}$, and set $g_1 = g_2 = g$ with $g/(2\pi) = \unit[40]{MHz}$ being an all-electrical magnon-photon coupling of similar magnitude as the one estimated in Eq.~\eqref{eq:electrical-g}.

\begin{figure}
    \centering
    \includegraphics[width=\columnwidth]{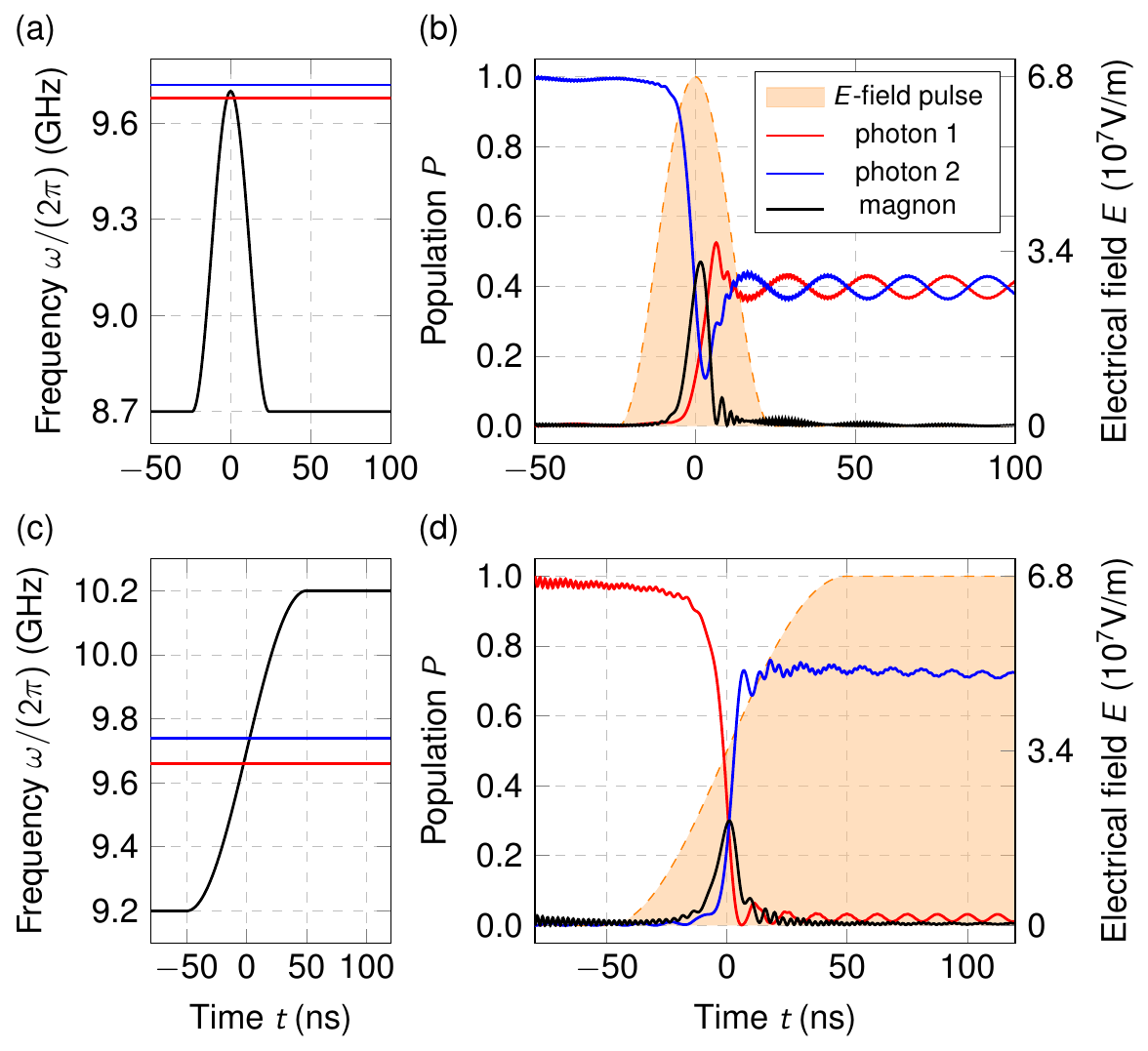}
    \caption{All-electrical coherent magnon-photon quantum 
    operations in the ferromagnetic phase of Cu$_2$OSeO$_3$. (a,b) Photon SPLIT operation with a full-period cosine $E$-field pulse. (a) Time dependence of the bare photon (red and blue) and magnon frequencies (black). (b) Time dependence of the modes' population upon the electric-field pulse. The quantum information of a maximally populated photon (blue) gets splitted into two parts (blue and red). (c,d) Photon SWAP operation with half-period cosine $E$-field pulse. (c) Time dependence of the bare photon (red and blue) and magnon frequencies (black). (d) Time dependence of the modes' population upon the electric-field pulse. The quantum information of a maximally populated (red) and empty photon (blue) gets swapped. For both operations, losses are associated with the magnon damping. For the parameters used, see text.}
    \label{fig:gatesFM}
\end{figure}

We consider two types of $E$-field protocols to electrically tune the magnon frequency in and out of resonance with the two photon frequencies. First, we concentrate on a full-period cosine protocol in the time interval $- \Delta t/2 < t \le \Delta t/2$, where
\begin{align}
    \omega_\text{m}(t) 
    = 
    \omega_\text{m}^{(0)} + \Delta \omega_\text{m} \frac{1-\cos( \omega_E (t+\Delta t/2) )}{2}.
\end{align}
Here, $\omega_\text{m}^{(0)}$ is the magnon frequency at zero electric field, $\Delta \omega_\text{m}/(2 \pi) = \unit[14.7]{Hz} \times 6.8 \times 10^{7}$ is the pulse amplitude, and $\omega_E = 2\pi / \Delta t$ denotes the pulse frequency. Figure~\ref{fig:gatesFM}(a) depicts how this protocol tunes the magnon frequency in and out of resonance with two photon frequencies $\Omega_1/(2\pi) = \unit[9.72]{GHz}$ and $\Omega_2/(2\pi) = \unit[9.68]{GHz}$, with $\omega_\text{m}^{(0)}/(2\pi) = \unit[8.7]{GHz}$ and $\omega_E/(2\pi) = \unit[20.83]{MHz}$. The maximal required field amplitude is $E = \unit[6.8 \times 10^7]{V/m}$, as shown in Fig.~\ref{fig:gatesFM}(b). Solving Eq.~\eqref{eq:EOM} numerically with this protocol, we trace the mode populations $P_1(t) = |a_1(t)|^2$, $P_2(t) = |a_2(t)|^2$, and $P_\text{m}(t) = |b(t)|^2$, with the initial conditions $P_1(0) = 0$, $P_2(0) = 1$, and $P_\text{m}(0) = 0$. As shown in Fig.~\ref{fig:gatesFM}(b), the field-protocol acts as a SPLIT operation, resulting in the transformation of the initial state into a superposition of output states.
Due to the damping of the magnon mode, the sum of final populations is smaller than the initial one.

As a second $E$-field protocol, we consider a half-period cosine pulse also in the time interval $- \Delta t/2 < t \le \Delta t/2$, where
\begin{align}
    \omega_\text{m}(t) 
    = 
    \omega_\text{m}^{(0)} + \Delta \omega_\text{m} \frac{1-\cos( \omega'_E (t+\Delta t/2) )}{2},
\end{align}
with $\omega'_E = 2\pi/(2\Delta t)$. For $t>\Delta t/2$, the electric field stays switched on, such that $\omega_\text{m}(t) = \omega_\text{m}^{(0)}+\Delta \omega_\text{m}$. This protocol is depicted in Fig.~\ref{fig:gatesFM}(c) for $\Omega_1/(2\pi) = \unit[9.74]{GHz}$ and $\Omega_2/(2\pi) = \unit[9.66]{GHz}$, $\omega_\text{m}^{(0)}/(2\pi) = \unit[9.2]{GHz}$, and $\omega'_E/(2\pi) = \unit[5]{MHz}$. The time-dependent mode populations shown in Fig.~\ref{fig:gatesFM}(d) reveal that the protocol realizes a SWAP operation (initial conditions: $P_1(0) = 1$, $P_2(0) = 0$, and $P_\text{m}(0) = 0$), with the quantum information getting swapped from photon 1 to photon 2.

\subsection{Photon-mediated coherent information transduction between magnons}
\label{sec:QI_application2}
Above, we concentrated on two-photon operations using magnons as a mediator. We now focus on the opposite idea of two-magnon operations using a photon as a mediator, as motivated by the recent interest in magnonic quantum states~\cite{yuanQuantumMagnonicsWhen2022} and facilitated by the magnetoelectric photon-mediated magnon-magnon coupling reported in Sec.~\ref{sec: SkX}.
In particular, we consider a SPLIT operation between the CCW and breathing modes. Similarly to Eq.~\eqref{eq:EOM}, the equation of motion is written as
\begin{align}
    \begin{pmatrix}
      \dot{a}(t) \\ \dot{b}_\text{B}(t) \\ \dot{b}_\text{C}(t)
    \end{pmatrix}
    =
    \begin{pmatrix}
        -\mathrm{i} \Omega - \frac{\kappa}{2} & -\mathrm{i} g_\text{B} & -\mathrm{i} g_\text{C} \\
        -\mathrm{i} g_\text{B}(t) & -\mathrm{i} \omega_\text{B}- \frac{\gamma_\text{B}}{2} & 0 \\
        -\mathrm{i} g_\text{C}(t) & 0 &  -\mathrm{i} \omega_\text{C} - \frac{\gamma_\text{C}}{2}
    \end{pmatrix}
    \begin{pmatrix}
      a(t) \\ b_\text{B}(t) \\ b_\text{C}(t)
    \end{pmatrix},
    \label{eq:EOM2}
\end{align}
where $\Omega$ is the photon frequency, $\omega_\text{B}(t) = \omega_\text{B}[E(t)]$ and $\omega_\text{C}(t) = \omega_\text{C}[E(t)]$ are the $E$-field dependent frequencies of the breathing and CCW modes, $\kappa$ is the photon damping, $\gamma_\text{B}$ and $\gamma_\text{C}$ are the damping of the breathing and CCW modes (full width at a half maximum), and $g_\text{B}$ and $g_\text{C}$ are the magnon-photon coupling of the breathing and CCW modes, respectively. 
We set $\Omega/(2\pi)=\unit[4.9]{GHz}$, $\kappa/(2\pi)=\unit[2]{MHz}$, and assume $\gamma_\text{B}=\gamma_\text{C}=0$ for simplicity. We note that the currently available estimates of the magnon damping in SkXs are too large for quantum gates between magnons~\cite{liensbergerTunableCooperativityCoupled2021a,khanCouplingMicrowavePhotons2021}. It is thus crucial to reduce the magnon damping, which could be realized in ultraclean materials at ultralow temperatures. In this context, we note that nonlinear magnon-photon coupling (e.g., two-magnon-one-photon interactions) might also give rise to additional channels of magnon damping, calling for a detailed theoretical analysis that we leave for further investigations. The results for zero magnon damping presented below demonstrate which operations could be achieved in principal.

\begin{figure}
    \centering
    \includegraphics[width=\columnwidth]{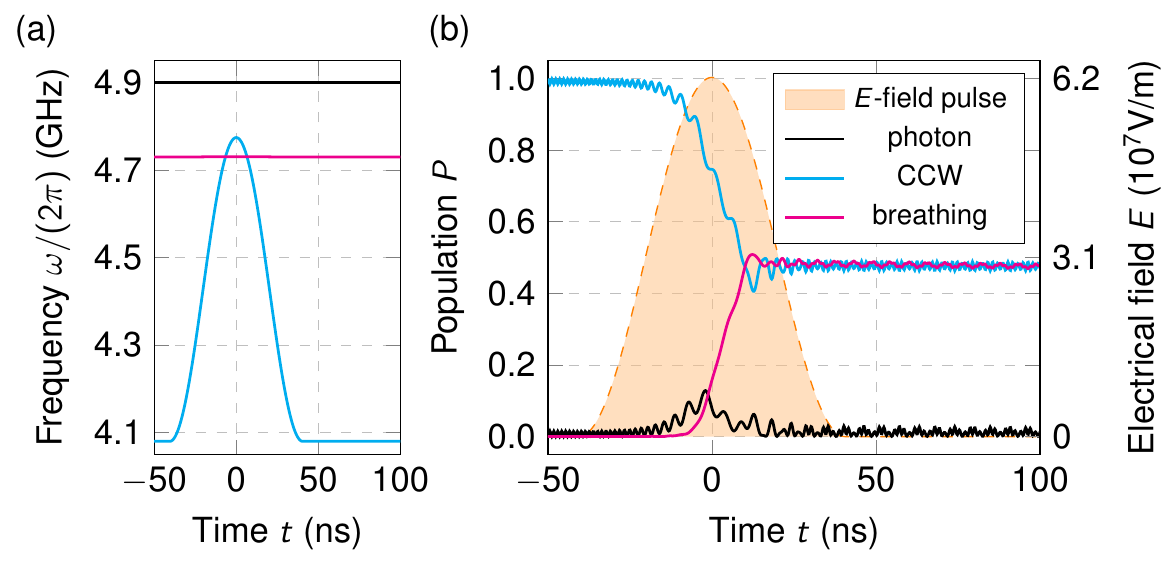}
    \caption{SPLIT operation of magnons mediated by photons in the SkX phase of Cu$_2$OSeO$_3$ under a full-period cosine $E$-field pulse. (a)~Time dependence of the bare photon frequency~(black) and bare frequencies of the CCW~(cyan) and breathing~(magenta) modes. (b) Time dependence of the modes' population upon the electric-field pulse. The quantum information of a maximally populated CCW mode~(cyan) gets splitted into two parts (cyan and magenta). Losses are associated with the photon damping; magnon damping is set to zero. For the parameters used, see text.}
    \label{fig:gatesSkX}
\end{figure}

Using the full-period cosine protocol, the $E$-field dependent magnon frequencies are defined as
\begin{subequations}
\label{eq: Time_EOM_SkX}
\begin{align}
    \omega_\text{B}(t) 
    &= 
    \omega_\text{B}^{(0)} + \Delta \omega_\text{B} \frac{1-\cos( \omega_E (t+\Delta t/2) )}{2},\\
    \omega_\text{C}(t) 
    &= 
    \omega_\text{C}^{(0)} + \Delta \omega_\text{C} \frac{1-\cos( \omega_E (t+\Delta t/2) )}{2},
\end{align}
\end{subequations}
with $- \Delta t/2 < t \le \Delta t/2$ and $\omega_E/(2\pi)=\unit[12.34]{MHz}$. We assume $B_z=\unit[112]{mT}$, where $\omega_\text{B}^{(0)}/(2\pi)=\unit[4.73]{GHz}$ and $\omega_\text{C}^{(0)}/(2\pi)=\unit[4.08]{GHz}$. Since the frequency shift of magnons in SkXs strongly depends on the direction of magnetization, applied electric fields, and the band index of magnon modes, it is possible to control the frequency of magnon modes selectively.~(For a detailed numerical investigation, see Appendix~\ref{sec: SkX_Stark}.) 
We set $\hat{z},-\vec{E}\parallel [111]$ with the maximal field amplitude $E=6.2\times \unit[10^7]{V/m}$ to achieve the largest frequency shift $\Delta \omega_\text{C}/(2\pi)=\unit[11.2]{Hz} \times 6.2\times 10^7$ in the CCW mode, while $\Delta \omega_\text{B}/(2\pi)=\unit[0.01]{Hz} \times 6.2\times 10^7$ in the breathing mode~[see Fig.\ref{fig:gatesSkX}(a)]. The coupling constants are obtained as $g_\text{B}/(2\pi)=\unit[24]{MHz}$ and $g_\text{C}/(2\pi)=\unit[52]{MHz}$ for $\vec{\CalB}\parallel[\overline{1}\overline{1}2]$, $\vec{\CalE}\parallel [111]$, $|\vec{\CalB}|=0.1\CalB_0$, and $|\vec{\CalE}|=0.995\CalE_0$~(cf. Appendix~\ref{section:coupling_strength}). With only the CCW mode initially occupied, i.e., $P(0) = 0$, $P_\text{B}(0) = 0$, and $P_\text{C}(0) = 1$, we solve Eq.~\eqref{eq:EOM2} numerically to study the time evolution of the modes' population. Figure~\ref{fig:gatesSkX}(b) clearly illustrates a SPLIT operation where the quantum information is split into both the CCW and breathing modes after the $E$-field pulse is applied.

\section{Conclusion} 
\label{sec: Conclusion}
We have studied the MP coupling in electromagnetic cavities hosting ME magnets, where a remarkable control over the coupling strength can be realized by rotating the magnetization direction. It even allows the strong enhancement of the MP coupling in one direction and the complete cancellation in the opposite direction. Focusing on the skyrmion-hosting multiferroic insulator Cu$_2$OSeO$_3$, we have shown that the strong coupling limit can be achieved electrically in the ferromagnetic phase. Furthermore, the ME coupling enables MP hybridization of the elliptic~(quadrupole) modes, which are ``magnetically dark'' excitations in the skyrmion crystal phase. We have also demonstrated that cavity modes can mediate the magnon-magnon coupling between the counterclockwise and breathing modes.
For the quantum information application of magnetoelectric magnon-photon coupling, we have proposed magnon-mediated quantum SPLIT and SWAP operations of photons in the collinear ferromagnetic phase, and a photon-mediated SPLIT operation of skyrmion eigenmodes, allowing all-electrical quantum transduction and entanglement generation.
Our theory also applies to other skyrmion-host materials, such as the lacunar spinels~\cite{kezsmarki2015neel,ruff2015multiferroicity}, by adjusting the dynamical electric moments $\vec{\pi}$ according to the relevant microscopic mechanism of spin-driven polarization~\cite{tokuraMultiferroicsSpinOrigin2014a}. We conclude that ME cavity magnonics provides an attractive platform for the coherent transduction between photons and magnons in topological magnets. 
The magnetoelectric MP hybridization and cavity-induced magnon-magnon coupling proposed in this work opens a novel path towards skyrmionic quantum-hybrid systems, quantum computing, and quantum magnonics.

\begin{acknowledgements}
    This work was supported by the Georg H. Endress Foundation and the Swiss National Science Foundation. This project received funding from the European Union's Horizon 2020 research and innovation program (ERC Starting Grant, Grant No.~757725).
\end{acknowledgements}



\appendix

\section{Magnetic multipole moments and electric dipole moments of magnons}
\label{sec:multipole_magnon}
In this section, we introduce macroscopic dynamical magnetic and electric multipole moments of magnons. The dimensionless magnetic dipole moment and quadrupole moment can be defined as~\cite{Schutte2014}
\begin{align}
M_a=\int_V \frac{d\vec{r}}{V} \,\hat{n}_a(\vec{r}), \,\,
Q_{ab}=
\int_V \frac{d\vec{r}}{V} \,\Big(\hat{n}_a(\vec{r})\hat{n}_b(\vec{r})-\frac{1}{3}\delta_{ab}(\vec{r})\Big),\label{eq:SM_MandQ}
\end{align}
respectively.
Here, $\hat{\vec{n}}(\vec{r})$ is a unit vector parallel to the magnetization and the integration is evaluated over the sample volume $V$. The subscript in $M_a$ refers to a component of $\vec{M}$ along a unit vector $\hat{\vec{a}}$. 
In order to compute multipole moments of magnons, we need to substitute $\hat{\vec{n}}(\vec{r},t)$ into Eq.~\eqref{eq:SM_MandQ} to find oscillations in multipole moments $M_a$ and $Q_{ab}$, where $\hat{\vec{n}}(\vec{r},t)$ represents the time evolution of spins when a magnon mode is excited~\cite{Diaz2020FM}. 

As an alternative approach, we provide a more convenient expression in terms of the wave function of the $n$th magnon mode. Replacing $\hat{\vec{n}}(\vec{r})$ with the dynamical dipole moment $\vec{\mu}_j$ ~[see Eq.~\eqref{eq:m1_expanded}], the magnonic dipole moment and quadrupole moment are respectively defined as
\begin{subequations}
\begin{align}
\CalM_a^{(n)}&= \frac{1}{\sqrt{N_\text{b}}} \sum_{ j=1 }^{N_\text{b}} \left(\mu_{j,a} u^{j,n}_{0} + \mu^*_{j,a} v^{j,n}_{0} \right),
\label{eq:SM_CalM}
\\
\CalQ^{(n)}_{ab}&= \frac{1}{\sqrt{N_\text{b}}} \sum_{ j=1 }^{N_\text{b}} \left(\mu_{j,a}\mu_{j,b} u^{j,n}_{0} + \mu^*_{j,a} \mu^*_{j,b}v^{j,n}_{0} \right),
\label{eq:SM_CalQ}
\end{align}
\end{subequations}
with $N_\text{b}$ denoting the number of spins in a magnetic unit cell. In Eqs.~\eqref{eq:SM_CalM} and~\eqref{eq:SM_CalQ}, the magnetic dipole moment and quadrupole moment are weighted by the wave function of the $n$th magnon mode, where a particle (hole) sector is represented by $u_{\vec{k}}^{j,n}$~($v_{\vec{k}}^{j,n}$) from Eq.~\eqref{eq:Bogo}. Similarly, we can define the electric dipole moment of the $n$th magnon mode as 
\begin{equation}
\CalP^{(n)}_a= \frac{1}{\sqrt{N_\text{b}}} \sum_{ j=1 }^{N_\text{b}} \left( \pi_{j,a} u^{j,n}_{0} + \pi^*_{j,a} v^{j,n}_{0} \right),\label{eq:SM_CalP}
\end{equation}
where the expressions for $\vec{\pi}_j$ is given in Eq.~\eqref{eq: dynamical_pi} of the main text for Cu$_2$OSeO$_3$. We note that the amplitude of multipole moments defined in Eqs.~\eqref{eq:SM_CalM}-\eqref{eq:SM_CalP} is independent of $N_\text{b}$, as shown for the skyrmion crystal~(SkX) in Appendix~\ref{sec:multipole_SkX}. This is understood from the normalized probability distribution of magnons, written as $\sum_{j=1}^{N_\text{b}}(|u_{0}^{j,n}|^2-|v_{0}^{j,n}|^2)=1$, which implies $u_{0}^{j,n},v_{0}^{j,n}\propto 1/\sqrt{N_\text{b}}$. As a result, the overall dependence in $N_\text{b}$ is canceled out in Eqs.~\eqref{eq:SM_CalM}-\eqref{eq:SM_CalP}.

Importantly, $\vec{\CalM}^{(n)}$ and $\vec{\CalP}^{(n)}$ naturally appear in the magnon-photon coupling as shown in Eq.~\eqref{eq:EMcoup}. Therefore, they represent macroscopic dynamical magnetic and electric dipole moments of the $n$th magnon mode, by which the latter couples to microwave fields. Furthermore, our expressions reveal that the coupling constant for magnon-photon interactions is enhanced with the total number of spins regardless of the size of magnetic unit cell. 

\section{Rescaling energies and magnetic fields for the low temperature skyrmion crystal phase of {Cu$_2$OSeO$_3$} }
\label{sec: rescaling}
To enhance the magnon-photon coupling, we consider the low-temperature SkX that can be realized in a thin film sample or by quenching a bulk sample~\cite{sekiObservationSkyrmionsMultiferroic2012,aqeelMicrowaveSpectroscopyLowTemperature2021b,takagiHybridizedMagnonModes2021}. Using the expressions for energies and magnetic fields in the continuum limit~\cite{Diaz2020FM}, we rescale the energy eigenvalues of magnons and magnetic fields as
\begin{subequations}
\begin{align}
    \tilde{\mathsf{E}}_n&=\frac{\tilde{J}}{\sqrt{3}}\left(\frac{\tilde{D}}{\tilde{J}}\right)^2\left(\frac{D}{J}\right)^{-2}\frac{\mathsf{E}_n}{JS}
    \label{eq:SM_tildeEn}
    ,\\
    \tilde{B}_z&=\frac{2\tilde{J}}{3}\left(\frac{\tilde{D}}{\tilde{J}}\right)^2\left(\frac{D}{J}\right)^{-2}\frac{\mathsf{g}\mu_BB_z}{JS},
    \label{eq:SM_tildeBz}
\end{align}
\end{subequations} with $D/J=0.5$ in our spin-lattice model and
$\tilde{D}/\tilde{J}=0.09$ in Cu$_2$OSeO$_3$~\cite{mochizukiDynamicalMagnetoelectricPhenomena2015}. The effective strength of ferromagnetic exchange interaction $\tilde{J}$ depends on temperatures and is estimated as $\tilde{J}=3$~meV at 5~K~\cite{mochizukiDynamicalMagnetoelectricPhenomena2015}. Here, we add numerical prefactors in Eqs.~\eqref{eq:SM_tildeEn} and~\eqref{eq:SM_tildeBz} to account for the triangular lattice. Substituting these values, we have $\tilde{\mathsf{E}}_n/h= \mathsf{E}_n/(JS)\times13.58$~GHz and $\tilde{B}_z=\mathsf{g} \mu_B B_z J / (D^2S)\times 140$~mT with the Planck constant~$h$.
In the main text, we drop tilde symbols to simplify the notation.

\section{Spin waves in skyrmion crystals}
\label{sec: spinwave_skx}
We compute the magnon spectrum of SkXs using linear spin-wave theory. Performing  the Holstein-Primakoff transformation as described in the main text, the bilinear magnon Hamiltonian is obtained by rewriting spin operators as $\vec{S}_i\approx \hat{\vec{e}}^z_i(S-b_i^\dagger b_i)+\sqrt{S/2}(\hat{\vec{e}}^-_ib_i+\hat{\vec{e}}^+_ib_i^\dagger)$, where $\hat{\vec{e}}_i^{\pm}=\hat{\vec{e}}_i^x\pm \mathrm{i} \hat{\vec{e}}_i^y$ and $\hat{\vec{e}}^z_i$ is a unit vector parallel to the classical ground-state spin configuration of SkXs. From Eq.~\eqref{eq: SpinLatticeH} of the main text, the linear spin-wave Hamiltonian reads up to a constant~\cite{Diaz2020FM,Hirosawa2020,hirosawaLasercontrolledRealReciprocalspace2021}:
\begin{equation}
    \mathcal{H}_\text{SW} =\frac{S}{2}\sum_{\vec{k},j,j'}\psi_{\vec{k},j}^\dagger 
    \begin{pmatrix}
    \Omega_{jj'}(\vec{k})& \Delta_{jj'}(\vec{k})\\
    \Delta_{jj'}^*(-\vec{k})& \Omega_{jj'}^*(-\vec{k})
    \end{pmatrix}
    \psi_{\vec{k},j'},
    \label{eq:SM_spinwave}
\end{equation}
where $\psi_{\vec{k},j}=(b_{\vec{k},j},b^\dagger_{-\vec{k},j})^\text{T}$. The Fourier transform of magnon operators is written as $b_{\vec{k},j}=\frac{1}{\sqrt{N_\text{u}}}\sum_i b_{i,j} \mathrm{e}^{-\mathrm{i}\vec{k}\cdot r_{i,j}} $, where $i$ and $j$ respectively represent a magnetic unit cell and sublattice site, and $N_\text{u}$ is the total number of magnetic unit cells.
We defined
\begin{subequations}
\begin{align}
    \Lambda_j(\vec{k})&=\sum_{j'} [J_{jj'}(\vec{k}=0)\hat{\vec{e}}^z_j\cdot\hat{\vec{e}}_{j'}^z-\vec{D}_{jj'}(\vec{k}=0)\cdot\hat{\vec{e}}^z_j\times \hat{\vec{e}}_{j'}^z]\nonumber \\
    &\quad +\frac{\mathsf{g}\mu_B B_z\hat{\vec{z}}\cdot\hat{\vec{e}}_j^z}{S},\\
    \Omega_{jj'}(\vec{k}) &=\delta_{jj'}\Lambda_j+\frac{1}{2}[-J_{jj'}(\vec{k})\hat{\vec{e}}_j^+\cdot \hat{\vec{e}}_{j'}^- +\vec{D}_{jj'}(\vec{k})\cdot \hat{\vec{e}}_j^+\times \hat{\vec{e}}_{j'}^-],\\
    \Delta_{jj'}(\vec{k})&=\frac{1}{2}[-J_{jj'}(\vec{k})\hat{\vec{e}}_j^+\cdot \hat{\vec{e}}_{j'}^+ +\vec{D}_{jj'}(\vec{k})\cdot \hat{\vec{e}}_j^+\times \hat{\vec{e}}_{j'}^+],
\end{align}
\end{subequations}
with $J_{jj'}(\vec{k})=\sum_i J_{\vec{r}_{i,j},\vec{r}_{0,j'}} \mathrm{e}^{-\mathrm{i}\vec{k}\cdot(\vec{r}_{i,j}-\vec{r}_{0,j'})}$ and $\vec{D}_{jj'}(\vec{k})=\sum_i \vec{D}_{\vec{r}_{i,j},\vec{r}_{0,j'}} \mathrm{e}^{-\mathrm{i}\vec{k}\cdot(\vec{r}_{i,j}-\vec{r}_{0,j'})}$. After diagonalization of Eq.~\eqref{eq:SM_spinwave}, we obtain 
\begin{equation}
    \mathcal{H}_\text{SW} =S\sum_{\vec{k},n}\mathsf{E}_n (\vec{k}) \left(b_{\vec{k},n}^\dagger b_{\vec{k},n}+\frac{1}{2} \right),
\end{equation}
where $\mathsf{E}_n (\vec{k})$ is the energy eigenvalue of $n$th magnon mode at crystal momentum $\vec{k}$ and $(b_{\vec{k},j},b_{-\vec{k},j})^T=T_{\vec{k}}^{j,n}(b_{\vec{k},n},b_{-\vec{k},n})^\text{T}$. We note that $T_{\vec{k}}^{j,n}$ is a paraunitary matrix~\cite{Colpa1978}, whose general expression is given in Eq.~\eqref{eq:Bogo}.

\begin{figure}[t]
    \centering
    \includegraphics[width=\columnwidth]{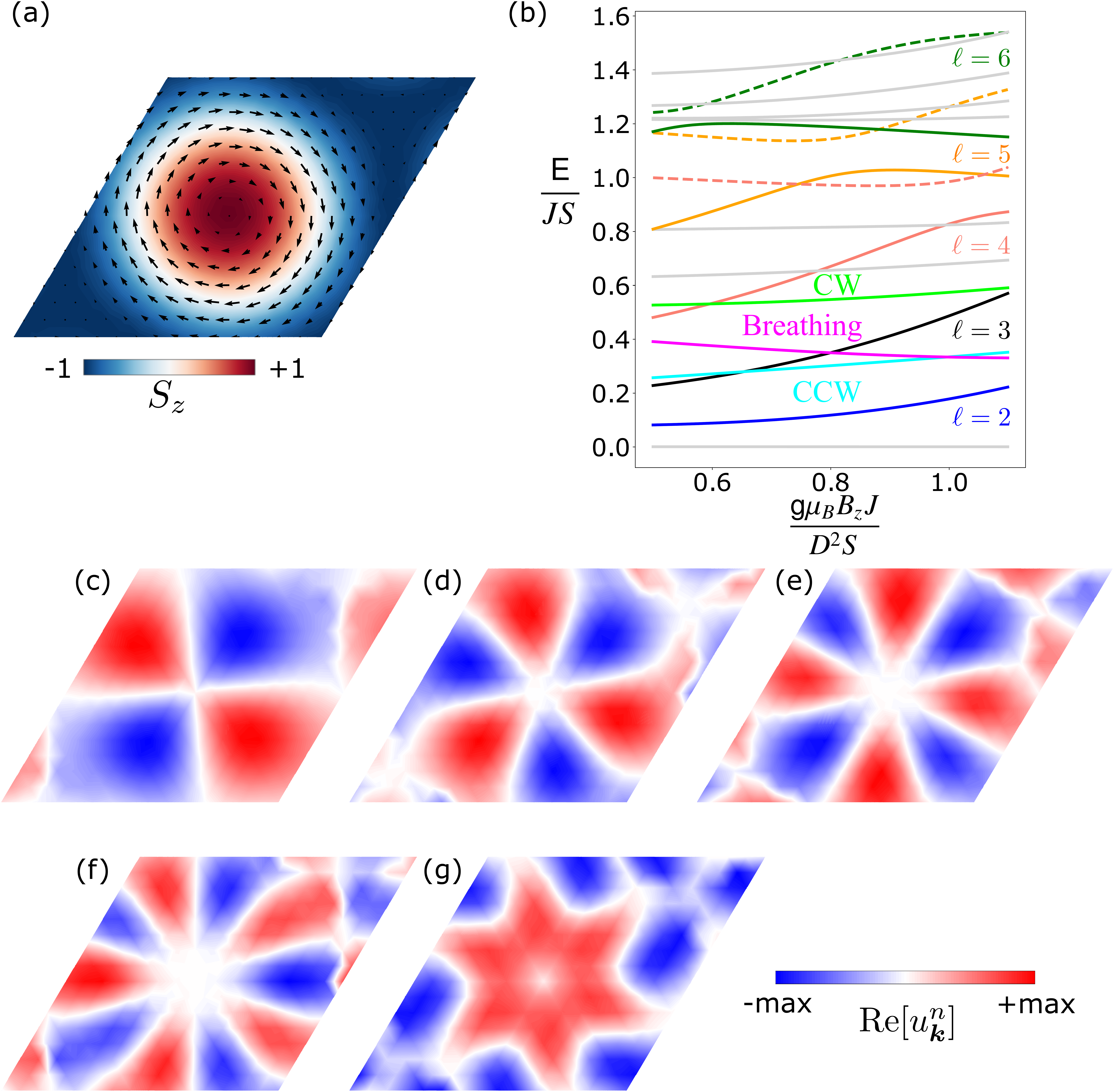}
    \caption{\textbf{Magnon spectrum of SkXs at the $\Gamma$ point.} (a) Classical ground-state spin configuration of SkXs, obtained at $D/J=0.5$ and $\mathsf{g} \mu_B B_z J / (D^2S)=0.55$ under periodic boundary conditions. The magnetic unit cell consists of $16\times 16$ spins on a triangular lattice. 
    (b)~Magnon spectrum of SkXs at $\vec{k}=0$ for the 18 lowest-energy eigenstates as a function of magnetic field. It highlights three magnetically active modes: CCW (cyan); breathing (magenta); CW (lime). In addition, the polygon modes are shown as $\ell=2$ (solid blue), $\ell=3$ (solid black), $\ell=4$ (solid salmon), $\ell=5$ (solid orange), and $\ell=6$ (solid green). Dashed lines indicate modes that hybridize with polygon modes for $\ell\ge 4$, corresponding to elliptic~(dashed salmon), CCW~(dashed orange), and breathing~(dashed green) modes. 
    (c-g) Wave function of polygon modes in SkXs.
    The color represents the real part of particle sector of $T_{\vec{k}}^{j,n}$, which is denoted as $u_{\vec{k}}^{j,n}$ in Eq.~\eqref{eq:Bogo}. Each panel corresponds to a polygon deformation mode with (c)~$\ell=2$, (d)~$\ell=3$, (e)~$\ell=4$, (f)~$\ell=5$, and (g)~$\ell=6$, respectively. 
    }
    \label{fig:SkX_band}
\end{figure}

Since the wavelength of microwave cavity modes is much larger than the size of the magnetic unit cell of SkXs, which is approximately 50~nm in Cu$_2$OSeO$_3$~\cite{sekiObservationSkyrmionsMultiferroic2012}, we only consider $\Gamma$ point magnon excitations in the following. Using the classical ground-state spin configuration of SkXs [cf. Fig.~\ref{fig:SkX_band}(a)], we obtain the magnon spectrum at $\vec{k}=0$ as a function of the static magnetic field $B_z$, as shown in Fig.~\ref{fig:SkX_band}(b).
Each magnon mode is identified by computing the time evolution of spin textures when the corresponding magnon mode is excited~\cite{Diaz2020FM}.
In addition to magnetically active modes, we highlight polygon deformation modes up to $\ell=6$. Interestingly, we notice that polygon modes with $\ell\ge 4$ show flattening or even decreasing energy eigenvalues as a function of magnetic fields in contrast to those hosted by single skyrmions~\cite{linInternalModesSkyrmion2014,Schutte2014}. 
Polygon modes are in fact strongly hybridized with an elliptic mode for $\ell=4$, CCW mode for $\ell=5$, and breathing mode for $\ell=6$, as indicated by large hybridization gaps between solid and dashed lines in Fig.~\ref{fig:SkX_band}(b). In contrast, the triangle mode is completely decoupled from CCW and breathing modes, showing band crossings. This is understood from the skyrmion-skyrmion interactions in SkXs, which breaks the rotational symmetry of polygon modes except for triangle modes. To support our argument, we discuss the hybridization of single-skyrmion bound states under perturbations in Appendix~\ref{sec:hybridizatin}. Figures~\ref{fig:SkX_band}(c-g) illustrate that the wave function of $\ell$th polygon mode has $C_\ell$ symmetry as expected. We note that the wave function for $\ell=6$ mode looks differently from other modes due to the hybridization with the second-harmonic breathing mode.

\begin{figure}[t]
    \centering
    \includegraphics[width=\columnwidth]{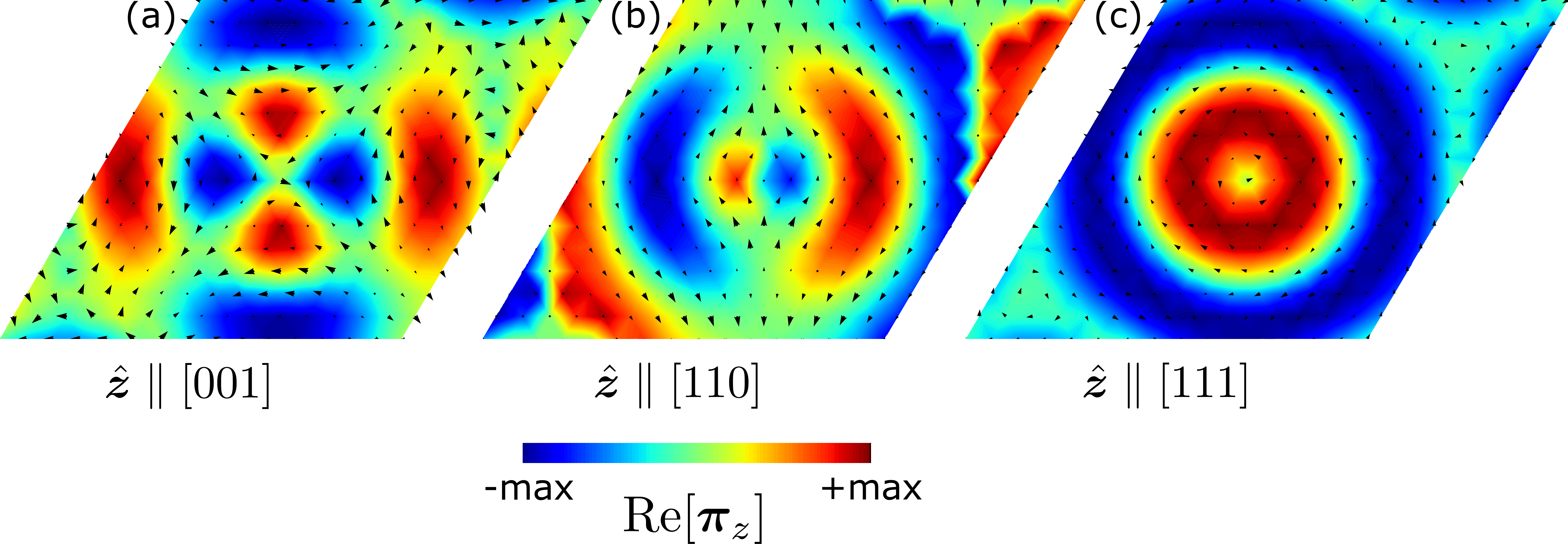}
    \caption{\textbf{Spatial profile of dynamical electric dipole moments in SkXs.} (a-c) Real part of $\vec{\pi}_i$ is plotted for the spin configuration of SkXs in Fig.~\ref{fig:SkX_band}(a) for (a)~$\hat{\vec{z}}\parallel[001]$, (b)~$\hat{\vec{z}}\parallel[110]$, and (c)~$\hat{\vec{z}}\parallel[111]$, which are defined in Eqs.~\eqref{eq:Pi_001}-\eqref{eq:Pi_111}.
    }
    \label{fig:SM_PiPlot}
\end{figure}

\section{Dynamical electric dipole moments in skyrmion crystals}
\label{sec:dynamical_SkX}
In the main text, we give the expression for $\vec{\pi}_i$ in terms of the crystallographic basis $(a,b,c)$ [see Eq.~\eqref{eq: dynamical_pi}]. While it is straightforward to apply the formula to the ferromagnetic phase, it is more convenient for the SkX phase to rewrite $\vec{\pi}_i$ in terms of another orthogonal basis $(x,y,z)$ whose out-of-plane component is antiparallel to the external magnetic field direction $\hat{\vec{z}}$. In the following, we consider three cases with $\hat{\vec{z}}\parallel [001], \, [110]\textrm{, and }[111]$ while $\hat{\vec{x}}\parallel [\overline{1}10]$ for all cases.

Firstly, we introduce a global rotation matrix that maps from $xyz$-coordinates to $abc$-coordinates as 
\begin{equation}
    \braket{\vec{S}_{i,abc}}=\hat{R}_{abc,xyz}\braket{\vec{S}_{i,xyz}}.
\end{equation}
Here, $
    \langle \vec{S}_{i,abc} \rangle 
    = 
    S
    (\cos \phi_i \sin \theta_i, \sin \phi_i \sin\theta_i,\cos\theta_i)^\text{T}   
$ and $
    \langle \vec{S}_{i,xyz} \rangle 
    = 
    S
    (\cos \phi'_i \sin \theta'_i, \sin \phi'_i \sin\theta'_i,\cos\theta'_i)^\text{T}   
$ respectively represent the classical ground-state spin configurations in $abc$- and $xyz$-coordinates.
We also need to define a local rotation matrix that maps from the local orthogonal basis ~$(\hat{\vec{e}}_i^x,\hat{\vec{e}}_i^y,\hat{\vec{e}}_i^z)$ to $xyz$-coordinates for the Holstein-Primakoff transformation, given as
\begin{equation}
R_{xyz,\hat{\vec{e}}_i}=
\begin{pmatrix}
\cos\theta'_i\cos\phi'_i&-\sin\phi'_i&\cos\phi'_i\sin\theta'_i\\
\cos\theta'_i\sin\phi'_i&\cos\phi'_i&\sin\theta'_i\sin\phi'_i\\
-\sin\theta'_i &0&\cos\theta'_i
\end{pmatrix}.
\end{equation}
Finally, we obtain the expression of $\vec{\pi}_i$ by collecting linear terms in magnon operators from the following expression:
\begin{align}
\vec{p}_{i,xyz}&=\hat{R}_{abc,xyz}^{-1}\,\vec{p}_{i,abc}\left(  \braket{\vec{S}_{i,abc}}\right)\nonumber\\
&=\hat{R}_{abc,xyz}^{-1}\,\vec{p}_{i,abc}\left(\hat{R}_{abc,xyz}\hat{R}_{\hat{xyz,\vec{e}}_i}\vec{S}_i\right),
\end{align}
where $\vec{S}_i\approx \hat{\vec{e}}^z_i(S-b_i^\dagger b_i)+\sqrt{S/2}(\hat{\vec{e}}^-_ib_i+\hat{\vec{e}}^+_ib_i^\dagger)$ and $\vec{p}_{i,abc}(  \braket{\vec{S}_{i,abc}})$ is given in Eq.~\eqref{eq:pCuOSeO} of the main text.

\begin{figure}[t]
    \centering
    \includegraphics[width=50mm]{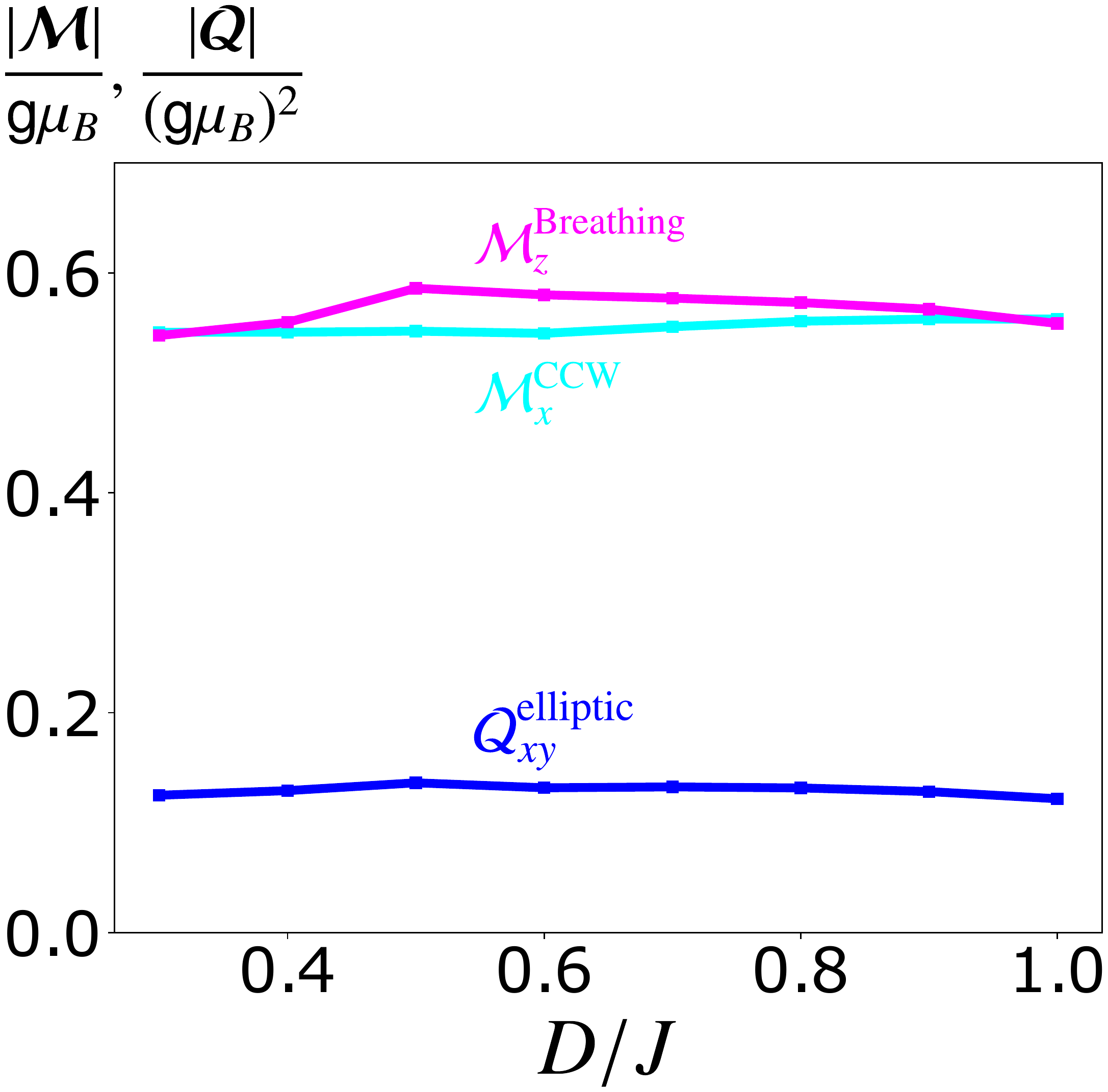}
    \caption{
    \textbf{Dynamical magnetic dipole moment $\vec{\CalM}$ and quadrupole moment $\vec{\CalQ}$ of magnons are independent of the size of skyrmions.} The dimensionless amplitudes of $\vec{\CalM}/(\mathsf{g}\mu_B)$ and $\vec{\CalQ}/(\mathsf{g}\mu_B)^2$ are plotted as a function of $D/J$. For each value of $D/J$ ratio, the classical ground-state spin configuration of SkXs is obtained at $\mathsf{g}\mu_BB_zJ/(D^2S)=0.8$, where $N_\text{b}=[576,361,256,169, 144,121,100,81]$ for $D/J=[0.3,0.4,0.5,0.6,0.7,0.8,0.9,1.0]$. 
    }
    \label{fig:MQ_DMI}
\end{figure}

\begin{figure}[t]
    \centering
    \includegraphics[width=0.95\columnwidth]{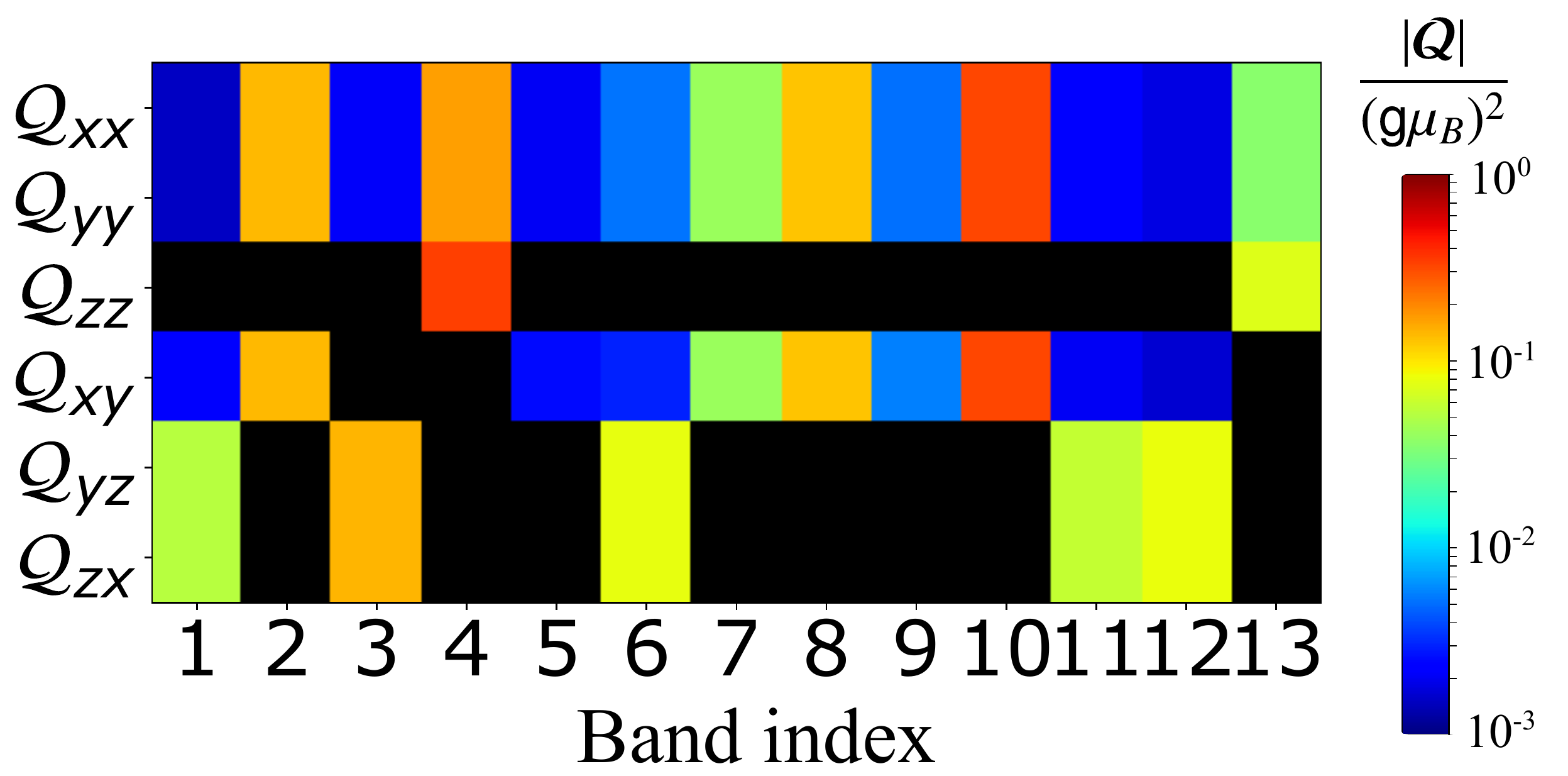}
    \caption{
    \textbf{Magnetic quadrupole moment $\CalQ$ of low-energy magnon modes in SkXs.} The dimensionless amplitude of $\CalQ/(\mathsf{g}\mu_B)^2$ is plotted on a logarithmic scale, obtained at $D/J=0.5$ and $\mathsf{g} \mu_B B_z J / (D^2S)=0.8$. Black regions indicate dipole moments smaller than $10^{-3}$. See Table~\ref{table: Multipole_expansion} for the description of magnon modes.
    }
    \label{fig:SM_quadrupole_SkX}
\end{figure}

\begin{table*}[t]
  \centering
   \caption{\textbf{Dynamical magnetic dipole moment, electric dipole moment, and magnetic quadrupole moment of low-energy magnon modes of SkXs in Cu$_2$OSeO$_3$.} For relative magnitudes, compare with Fig.~\ref{fig: dipole} in the main text and Fig.~\ref{fig:SM_quadrupole_SkX}. If the corresponding single-skyrmion bound states are magnetically~(electrically) active, they are highlighted in bold red~(blue). In the ``Index'' column, the colors in brackets coincide with the colors chosen in Fig.~\ref{fig:SkX_band}(b). We denote $\boldsymbol{\CalP}^{\alpha},\, \boldsymbol{\CalP}^{\beta}, \textrm{ and }\boldsymbol{\CalP}^{\gamma}$ as electric polarization with $\hat{\vec{z}}\parallel [001]$, $[110]$, and $[111]$, respectively. For all cases, the $x$-axis is taken along $[\overline{1}10]$.
   }
 \label{table: Multipole_expansion}
  \begin{tabular}{l c c c c}
  \toprule
  Index&Description&$\boldsymbol{\CalM}$ &$\boldsymbol{\CalP}$ & $\boldsymbol{\CalQ}$ \\ \cline{1-5} 
  1 (solid grey) & Goldstone mode & $-$ & $-$ &$\CalQ_{yz}, \CalQ_{zx}$ \\
  2 (solid blue) & \textbf{\textcolor{blue}{Elliptic}}  & $-$ & $\CalP_z^{\alpha}, \CalP_x^{\beta}, \CalP_y^{\beta}, \CalP_x^{\gamma}, \CalP_y^{\gamma} $ & $\CalQ_{xx}, \CalQ_{yy},\CalQ_{xy}$ \\  
  3 (solid cyan) & \textbf{\textcolor{red}{CCW}} &  $\CalM_x, \CalM_y$&$\CalP_x^{\alpha},\CalP_y^{\alpha},\CalP_z^{\beta},\CalP_x^{\gamma},\CalP_y^{\gamma}$ &$\CalQ_{yz}, \CalQ_{zx}$  \\
  4 (solid magenta) & \textbf{\textcolor{red}{Breathing}} & $\CalM_z$& $\CalP_y^{\beta}, \CalP_z^{\gamma}$ & $\CalQ_{xx}, \CalQ_{yy},\CalQ_{zz}$ \\
  5 (solid black) & Triangle & $-$ & $-$ & $-$ \\
  6 (solid lime) & \textbf{\textcolor{red}{CW}} & $\CalM_x, \CalM_y $& $\CalP_x^{\alpha},\CalP_y^{\alpha}, \CalP_z^{\beta}, \CalP_x^{\gamma},\CalP_y^{\gamma} $ &$\CalQ_{yz}, \CalQ_{zx}$\\
  7 (solid grey) & \textbf{\textcolor{blue}{Elliptic}}  & $-$ & $\CalP_z^{\alpha}, \CalP_x^{\beta}, \CalP_y^{\beta}, \CalP_x^{\gamma}, \CalP_y^{\gamma}$ &$\CalQ_{xx}, \CalQ_{yy},\CalQ_{xy}$ \\
  8 (solid salmon) & Square  & $-$ &$\CalP_z^{\alpha}, \CalP_x^{\beta}, \CalP_y^{\beta}, \CalP_x^{\gamma},\CalP_y^{\gamma}$ & $\CalQ_{xx}, \CalQ_{yy},\CalQ_{xy}$\\
  9 (solid grey) & Triangle  & $-$ & $-$ & $-$ \\
  10 (dashed salmon) & \textbf{\textcolor{blue}{Elliptic}} & $-$  & $\CalP_z^{\alpha},\CalP_x^{\beta}, \CalP_y^{\beta}, \CalP_x^{\gamma}, \CalP_y^{\gamma}$ &$\CalQ_{xx}, \CalQ_{yy},\CalQ_{xy}$ \\
  11 (solid orange) & Pentagon  & $ \CalM_x, \CalM_y$& $\CalP_x^{\alpha}, \CalP_y^{\alpha}, \CalP_z^{\beta}, \CalP_x^{\gamma}, \CalP_y^{\gamma}$ & $\CalQ_{yz}, \CalQ_{zx}$\\
  12 (dashed orange) & \textbf{\textcolor{red}{CCW}}&  $ \CalM_x, \CalM_y$ & $\CalP_x^{\alpha}, \CalP_y^{\alpha}, \CalP_z^{\beta}, \CalP_x^{\gamma},\CalP_y^{\gamma}$ & $\CalQ_{yz}, \CalQ_{zx}$\\
  13 (solid green) & Hexagon & $\CalM_z$  & $\CalP_y^{\beta}, \CalP_z^{\gamma}$ & $\CalQ_{xx}, \CalQ_{yy},\CalQ_{zz}$\\
  \botrule
 \end{tabular}
\end{table*}

The expressions for $\vec{\pi}_i$ are respectively obtained as
\begin{align}
\vec{\pi}_i&=\frac{\lambda S^{\frac{3}{2}}}{\sqrt{2}}
\begin{pmatrix}
-\cos\phi_i'\cos2\theta_i'-\mathrm{i}\sin\phi_i'\cos\theta_i'\\
\sin\phi_i'\cos2\theta_i'-\mathrm{i}\cos\phi_i'\cos\theta_i'\\
-\cos2\phi_i'\sin\theta_i'\cos\theta_i'-\mathrm{i}\sin2\phi_i'\sin\theta_i'
\end{pmatrix}\nonumber\\
&\textrm{ for }\hat{\vec{z}}\parallel [001],
\label{eq:Pi_001}
\end{align}
\begin{align}
\vec{\pi}_i&=\frac{\lambda S^{\frac{3}{2}}}{\sqrt{2}}
\begin{pmatrix}
(-\sin2\phi_i'\cos\theta_i'+\mathrm{i}\cos2\phi_i')\sin\theta_i'\\
-\frac{1}{4}(3+\cos2\phi_i')\sin2\theta_i'-\mathrm{i}\cos\phi_i'\sin\phi_i'\sin\theta_i'\\ \cos2\theta_i'\sin\phi_i'-\mathrm{i}\cos\phi_i'\cos\theta_i'
\end{pmatrix}\nonumber\\
&\textrm{ for }\hat{\vec{z}}\parallel [110],
\label{eq:Pi_110}
\end{align}
and
\begin{align}
\boldsymbol{\pi}_i&=\frac{\lambda S^{\frac{3}{2}}}{2\sqrt{3}}
\left(\begin{matrix}
\sqrt{2}(-\cos\phi_i'\cos2\theta_i'-\mathrm{i}\cos\theta_i'\sin\phi_i')\\
\sqrt{2}(-\sin\phi_i'\cos2\theta_i'+\mathrm{i}\cos\theta_i'\cos\phi_i')\\
-3\sqrt{2}\cos\theta_i'\sin\theta_i'
\end{matrix}\right.
\nonumber\\
&\quad\quad\quad\quad\quad\quad
\left.\begin{matrix}
-\sin2\phi_i'\sin2\theta_i'+2\mathrm{i}\cos2\phi_i'\sin\theta_i',\\
-\cos2\phi_i'\sin2\theta_i'-2\mathrm{i}\sin2\phi_i'\sin\theta_i'\\
\\
\end{matrix}\right)
\nonumber\\
&\textrm{ for }\hat{\vec{z}}\parallel [111].
\label{eq:Pi_111}
\end{align}

We note that the symmetry of $\vec{\pi}_i$ is consistent with the magnetic point group for each case. For example, we find that $\vec{\pi}_i\rightarrow C_{2z}(\vec{\pi}_i)$ by substituting $\phi'_i\rightarrow \phi'_i+\pi$ for $\hat{\vec{z}}\parallel [001]$. This implies that $\vec{\pi}_i$ is symmetric about $C_{2z}$ rotation~\cite{hirosawaLasercontrolledRealReciprocalspace2021}, which is a symmetry of the corresponding magnetic point group~$2'2'2$ for $\hat{\vec{z}}\parallel [001]$~\cite{mochizukiDynamicalMagnetoelectricPhenomena2015}. Similarly, we find that $\vec{\pi}_i$ is symmetric about $C_{2y}\mathcal{T}$~($C_{3z}$) for $\hat{\vec{z}}\parallel[110]$~($\hat{\vec{z}}\parallel[111]$). This is demonstrated in Fig.~\ref{fig:SM_PiPlot}, showing spatial profiles of the real part of $\vec{\pi}_i$ obtained from the spin configuration of SkXs in Fig.~\ref{fig:SkX_band}(a). It is important to note that the out-of-plane component of dynamical electric dipole moment $\vec{\pi}$  for $\hat{\vec{z}}\parallel [001]$ shows a quadrupolar feature similarly to the ground-state electric polarization $\vec{P}$~(see Fig.~\ref{fig:teaser} of the main text). Therefore, the electric coupling allows us to study the quadrupole moment of magnons, which is defined in Appendix~\ref{sec:multipole_magnon}.

\section{Magnetic multipole moments and electric dipole moments in skyrmion crystals}
\label{sec:multipole_SkX}
In this section, we discuss magnetic and electric multipole moments of low-energy magnon modes in skyrmion crystals. Firstly, we demonstrate that the amplitude of multipole moments defined in Eqs.~\eqref{eq:SM_CalM}-\eqref{eq:SM_CalP} is independent of $N_\text{b}$. This is illustrated in Fig.~\ref{fig:MQ_DMI}, where we investigate $\CalM_x^\textrm{CCW}$, $\CalM_z^\textrm{Breathing}$, and $\CalQ_{xy}^\textrm{Elliptic}$ at different $D/J$ ratios. Since the linear size of skyrmions is approximately proportional to $(D/J)^{-1}$, the total number of spins in each magnetic unit cell scales as $N_\text{b} \propto (D/J)^{-2}$. 
Figure~\ref{fig:MQ_DMI} clearly demonstrates that the amplitudes of $\vec{\CalM}$ and $\vec{\CalQ}$ are constant for various sizes of skyrmions stabilized at different $D/J$ ratios. We have also confirmed that $\vec{\CalP}$ is independent of $D/J$ ratios, where the expressions for $\vec{\pi}_j$ are given in Eqs.~\eqref{eq:Pi_001}-\eqref{eq:Pi_111}. Magnetic dipole moments, electric dipole moments, and magnetic quadrupole moments of low-energy magnon modes in SkXs are summarized in Fig.~\ref{fig:SM_quadrupole_SkX} and Table~\ref{table: Multipole_expansion}. We note that our definition of $\CalQ_{ab}$ is consistent with the previous work, where $Q_{ab}$ in Eq.~\eqref{eq:SM_MandQ} was computed for the elliptic modes and breathing modes~\cite{Schutte2014}.

\section{Hybridization of single-skyrmion bound states}
\label{sec:hybridizatin}
In the main text, we show that all polygon deformation modes except for the triangle modes carry nonzero $\vec{\CalM}$ and/or $\vec{\CalP}$ in SkXs, although single-skyrmion bound states, characterized by azimuthal quantum number~$\ell$~\cite{shekaInternalModesMagnon2001}, cannot have magnetic/electric dipole moments except for the elliptic modes besides CCW, CW, and breathing modes~\cite{Schutte2014}.
In Appendix~\ref{sec: spinwave_skx}, we demonstrate that this is because the skyrmion-skyrmion interactions strongly hybridize single-skyrmion bound states with different values of $\ell$ in SkXs. Here, we provide a pedagogical example by studying single-skyrmion bound states under symmetry-breaking perturbations.

\begin{figure}[t]
    \centering
    \includegraphics[width=\columnwidth]{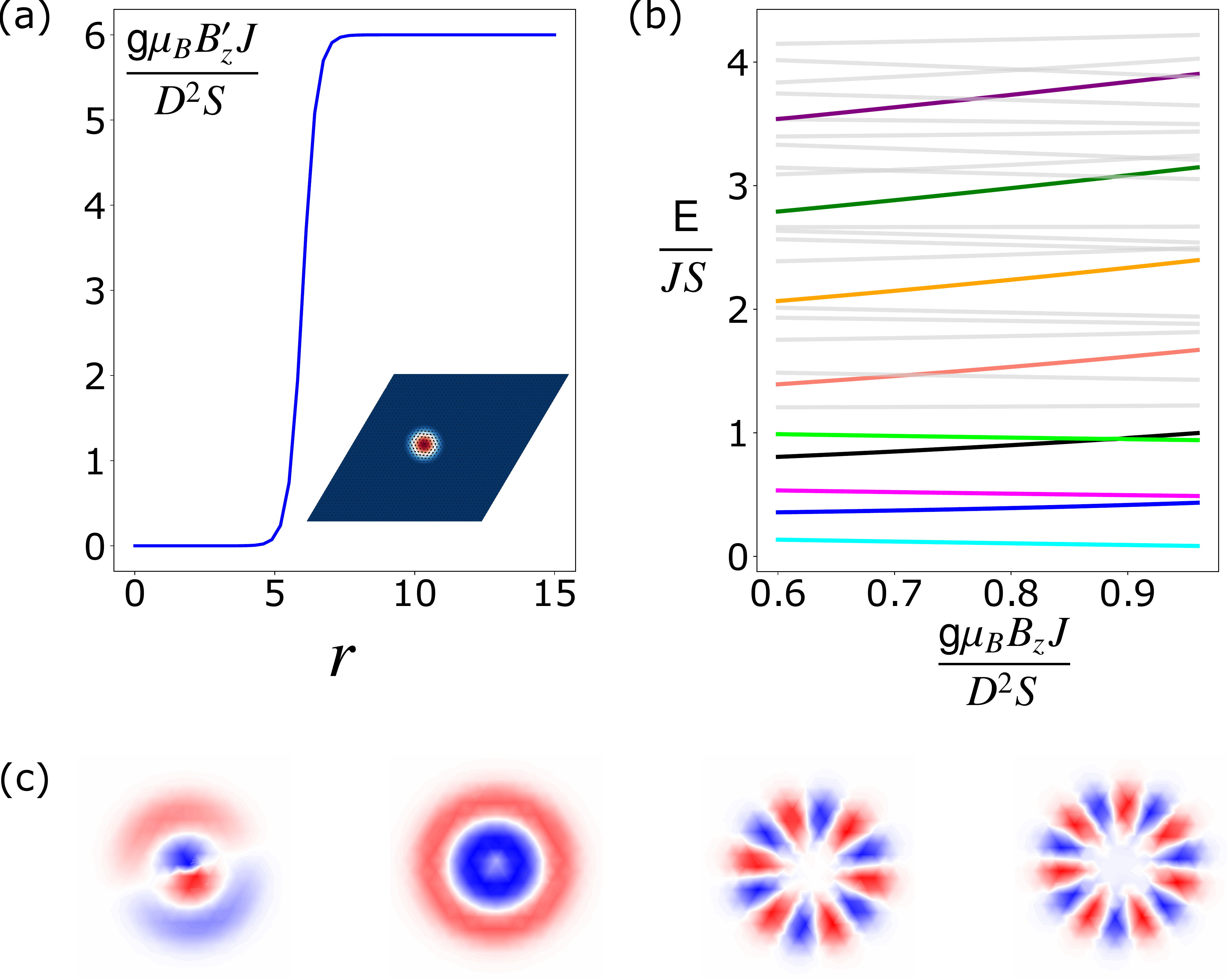}
    \caption{
    \textbf{Single-skyrmion bound states trapped inside the magnetic field potential.} (a)~Spatial profile of magnetic field $B'_z$ from the center of the skyrmion, added on top of uniform magnetic field $B_z$. The expression for $B'_z$ is given in Eq.~\eqref{eq:SM_radialB}. The inset shows the spin configuration of a single skyrmion inside the magnetic field trap, obtained at $D/J=0.5$ for $50\times 50$ spins with periodic boundaries. (b)~Spectrum of single-skyrmion bound states at $\vec{k}=0$ as a function of uniform magnetic field $B_z$. Highlighted modes are CCW~(cyan), breathing~(magenta), CW~(lime), $\ell=2$~(blue), $\ell=3$~(black), $\ell=4$~(salmon), $\ell=5$~(orange), $\ell=6$~(green), and $\ell=7 $~(purple). 
    (c)~Wave function of high-energy skyrmion bound states obtained at $\mathsf{g} \mu_B B_z J / (D^2S)=0.65$. From left to right, we show second-order CCW, second-order breathing modes, $\ell=6$, and $\ell=7$. Their eigenenegies are $\mathsf{E}/(JS)=1.21,~ 1.92,~2.84,$ and 3.58. 
    The color code is the same as in Fig.~\ref{fig:SkX_band}(c-g). 
    }
    \label{fig:SM_singleSk}
\end{figure}

According to previous studies, most of the bound states of a single skyrmion are found above the bulk continuum excitations of the ferromagnetic background~\cite{linInternalModesSkyrmion2014,Schutte2014,diazSpinWaveRadiation2020a}. To eliminate the bulk continuum from the low-energy spectrum, we introduce a magnetic field potential trap 
[see Fig.~\ref{fig:SM_singleSk}(a)]~\cite{wangTheorySkyrmionSize2018}: 
\begin{align}
    &\frac{\mathsf{g}\mu_B B'_z(\vec{r})J}{D^2S}
    =
    6\big(\cos\theta(\vec{r})+1\big),\nonumber\\
    &\theta(\vec{r})
    =
    2\arctan\left(\frac{\sinh (R/\Delta)}{\sinh (r/\Delta)}\right),
    \label{eq:SM_radialB}
\end{align}
with $\vec{r}$ denoting displacement from the skyrmion center, $R=6$ for skyrmion radius, and $\Delta=0.5$. The classical ground-state spin configuration is prepared by Monte Carlo annealing to create an isolated single skyrmion at $D/J=0.5$ and $\mathsf{g}\mu_BB_zJ/(D^2S)=1.0$. The obtained spin configuration is then relaxed under the additional magnetic field potential $B_z'$ with LLG simulations.  

In the presence of a large magnetic field potential $B'_z$, many single-skyrmion bound states are successfully obtained in the magnon spectrum as shown in Fig.~\ref{fig:SM_singleSk}(b). While the elliptic mode is known to cause a bimeron instability below the critical magnetic field~\cite{ezawaCompactMeronsSkyrmions2011}, the magnetic field potential prevents such instability. Hence, the lowest-energy mode is found to be the CCW mode instead of the elliptic mode. At higher energies, we find other polygon deformation modes that are highlighted in Fig.~\ref{fig:SM_singleSk}(b). In addition, we find second-order and even higher-order harmonics, as shown in Fig.~\ref{fig:SM_singleSk}(c). In contrast to the magnon spectrum of SkXs, there is no signature of avoided crossings in Fig.~\ref{fig:SM_singleSk}(b). Analogously to atomic orbitals of hydrogen atoms, the azimuthal quantum number $\ell$ is a conserved quantity of single-skyrmion bound states, thus preventing hybridization between different $\ell$ states. We also confirm that only the elliptic modes carry finite $\vec{\CalP}$ in addition to CCW, CW, and breathing modes.

\begin{figure}[t]
    \centering
    \includegraphics[width=\columnwidth]{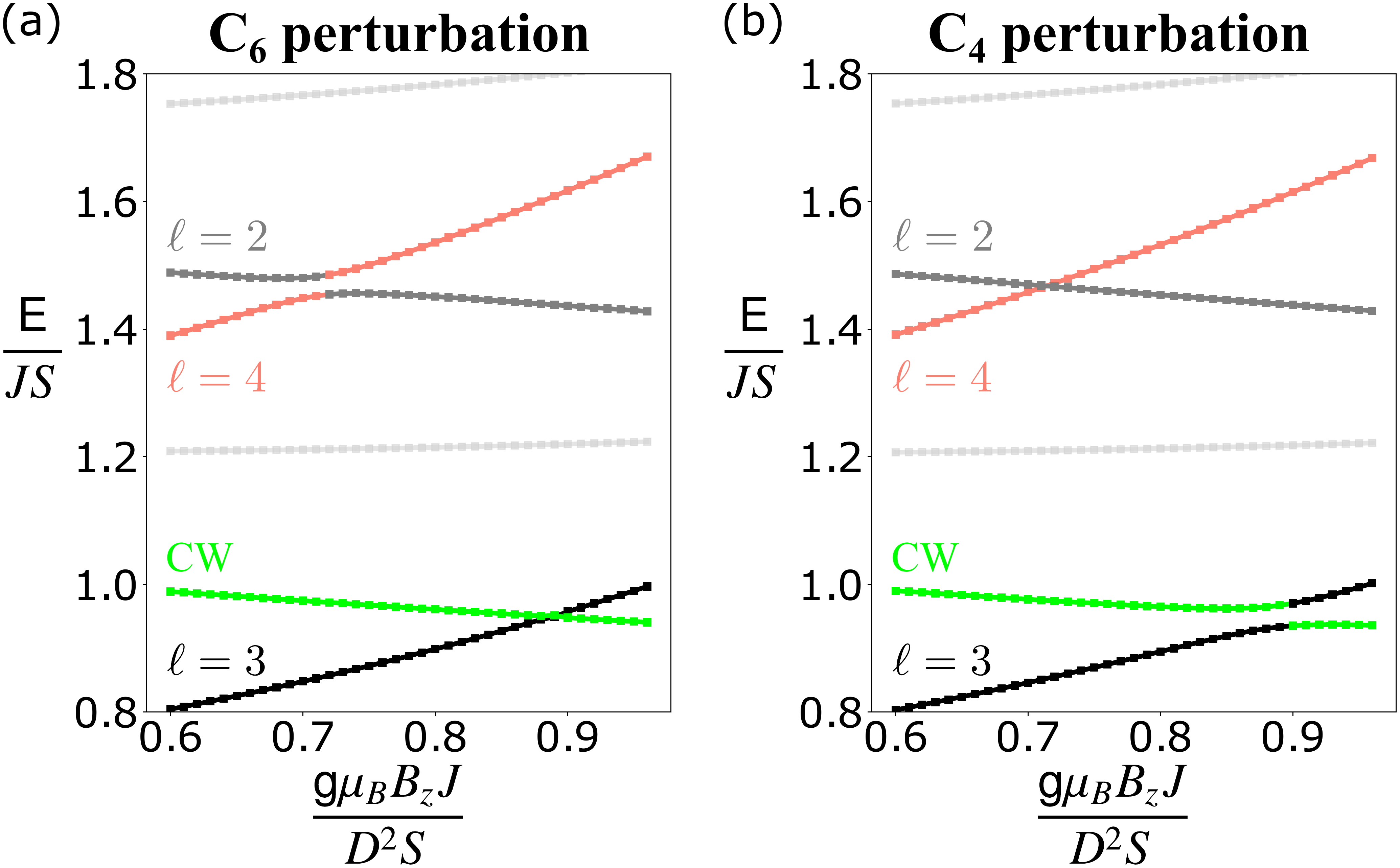}
    \caption{
    \textbf{Hybridization of single-skyrmion bound states under symmetry-breaking perturbations.} (a,b) Magnon spectrum of a single skymion at $\vec{k}=0$ under $B_z+B_z'+\delta B_z$, where $B'_z$ and $\delta B_z$ are respectively defined in Eqs.~\eqref{eq:SM_radialB} and~\eqref{eq:SM_perturb}. The continuous rotational symmetry of skyrmions is broken by (a) $C_6$ symmetric perturbations and (b) $C_4$ symmetric perturbations. The symmetry-breaking perturbation results in hybridization between the lowest $\ell=4$ mode~(salmon) and second $\ell=2$ mode~(dark grey) in~(a), and between the lowest $\ell=3$ mode~(black) and CW mode~(lime) in~(b).
    }
    \label{fig:SM_hybridization}
\end{figure}

Now, we break the continuous rotational symmetry by adding small perturbations of the form
\begin{equation}
    \frac{\mathsf{g}\mu_B\delta B_z(\vec{r})J}{D^2S}=0.04 \cos \left[q\phi(\vec{r})  \right]H_s(R-r),
    \label{eq:SM_perturb}
\end{equation}
with $\phi$ denoting an azimuthal angle from the center of the skyrmion, integers $q$ for $C_q$ symmetric perturbations, and $H_s(x)$ representing the Heaviside step function. In the following, we consider perturbations with $q=6$ and $q=4$ to simulate the skyrmion-skyrmion interactions of hexagonal and square SkXs. For each case, the classical spin configuration of a single skyrmion is relaxed under $B_z+B_z'+\delta B_z$, which is then used to compute the magnon spectrum. Figure~\ref{fig:SM_hybridization} shows magnon spectra (a) for $q=6$ and (b) $q=4$. It is clearly demonstrated that the $C_6$ symmetric perturbation results in hybridization of the lowest square mode~(salmon) with the second elliptic mode~(dark grey), while the $C_4$ symmetric perturbation hybridizes the triangle mode~(black) with the CW mode~(lime).  
This is because the azimuthal quantum number is no longer a good quantum number except for $\ell=\frac{q}{2}$ polygon modes, noting that $\ell$th order polygon modes carry multipole moment of order $2\ell$. Therefore, only an $\ell=\frac{q}{2}$ polygon mode remains an exact eigenstate, rendering the triangle modes elusive for experimental observations in hexagonal SkXs.

\section{Magnon-photon coupling in skyrmion crystals}
\label{section:coupling_strength}
In this section, we estimate the magnon-photon coupling strength in SkXs. For simplicity, we consider low-energy magnon modes of SkXs coupled with a single cavity mode.
Using the rotating wave approximation, the effective magnon-photon coupling is given as 
\begin{equation}
\mathcal{H}_\text{EM} =\sum_{n=1}^{18}\left( \hbar g^{(n)}a^\dagger b_{0,n} +\hbar \left(g^{(n)}\right)^*a b^\dagger_{0,n}\right),
\label{eq:SM_hybrid}
\end{equation}
where magnon modes are included up to $n=18$, corresponding to those shown in Fig.~\ref{fig:SkX_band}(b). The coupling constant is given as~[see Eq.~\eqref{eq:EMcoup}]
\begin{equation}
    \hbar g^{(n)}=-\sqrt{N}(\vec{\CalE}^*\cdot \vec{\CalP}^{(n)}+\vec{\CalB}^*\cdot \vec{\CalM}^{(n)}),
    \label{eq:SM_gvalue}
\end{equation}
where $\vec{\CalE}$ and $\vec{\CalB}$ are quantized electromagnetic fields at the position of the SkX.
We note that Eq.~\eqref{eq:SM_hybrid} is derived as a perturbation, assuming that the coupling constant is much smaller than the energy eigenvalues of magnon and cavity modes.

The maximum amplitude of electromagnetic fields inside a cavity is given as $\CalB_0=\sqrt{\hbar \Omega\mu_0/(2V)}$ and $\CalE_0=\sqrt{\hbar \Omega/(2\epsilon_0 V)}$~\cite{Huebl2013}, where $\mu_0$ and $\epsilon_0$ respectively denote the vacuum permeability and vacuum permittivity, $\Omega$ is the frequency of microwave cavity, and $V$ is the volume of cavity. Substituting into Eq.~\eqref{eq:SM_gvalue}, the maximum magnetic and electric components of MP coupling are given as
\begin{align}
    \hbar g^{(n)}_\textrm{m-mag}
     &=-\sqrt{\frac{\hbar\Omega \mu_0\rho_s\eta}{2}}\,\hat{\vec{\CalB}}\cdot \vec{\CalM}^{(n)},
     \label{eq:SM_g_mag}
     \\
     \hbar g^{(n)}_\textrm{m-el}
     &=-\sqrt{\frac{\hbar\Omega \rho_s\eta}{2\epsilon_0}}\,\hat{\vec{\CalE}}\cdot \vec{\CalP}^{(n)}\label{eq:SM_g_el}
\end{align}
with $\rho_s\approx 5.67\times 10^{27}$~m$^{-3}$ being the spin density of Cu$_2$OSeO$_3$~\cite{mochizukiDynamicalMagnetoelectricPhenomena2015} and $\eta=V_\textrm{crys}/V$ the volume ratio between the crystal and cavity. Here, $\hat{\vec{\CalB}}~(\hat{\vec{\CalE}})$ represents a unit
vector parallel to magnetic (electric) fields of the cavity mode.
In order to realize a strong coupling between magnon and cavity modes, it is crucial to increase the value of $\eta$. However, it also needs to be sufficiently small to have an approximately spatially uniform microwave field inside the sample.
In the previous experiments, the parameters were chosen as $\eta\approx 0.03$ and $\Omega/(2\pi)=1.1$~GHz in Ref.~\cite{khanCouplingMicrowavePhotons2021} and $\eta\approx 0.06$ and $\Omega/(2\pi)=0.68$~GHz in Ref.~\cite{liensbergerTunableCooperativityCoupled2021a}, where $|g_\textrm{m-mag}^\textrm{CCW}|/(2\pi)$ was found to be approximately 120~MHz in both references. Substituting $|\vec{\CalM}_{x}^\textrm{CCW}|=0.55\mathsf{g}\mu_B$~(see Fig.~\ref{fig:MQ_DMI}), we obtain $|g_\textrm{m-mag}^\textrm{CCW}|/(2\pi)=135$~MHz and~150~MHz, respectively. Hence, our theory is in good agreement with the experiments. In the following, to enhance the magnon-photon coupling, we use the rescaled magnon energies~$\tilde{\mathsf{E}}_{n}$ and magnetic fields~$\tilde{B}_z$ for the low temperature SkX phase provided in Appendix~\ref{sec: rescaling}.

\begin{figure}[t]
    \centering
    \includegraphics[width=\columnwidth]{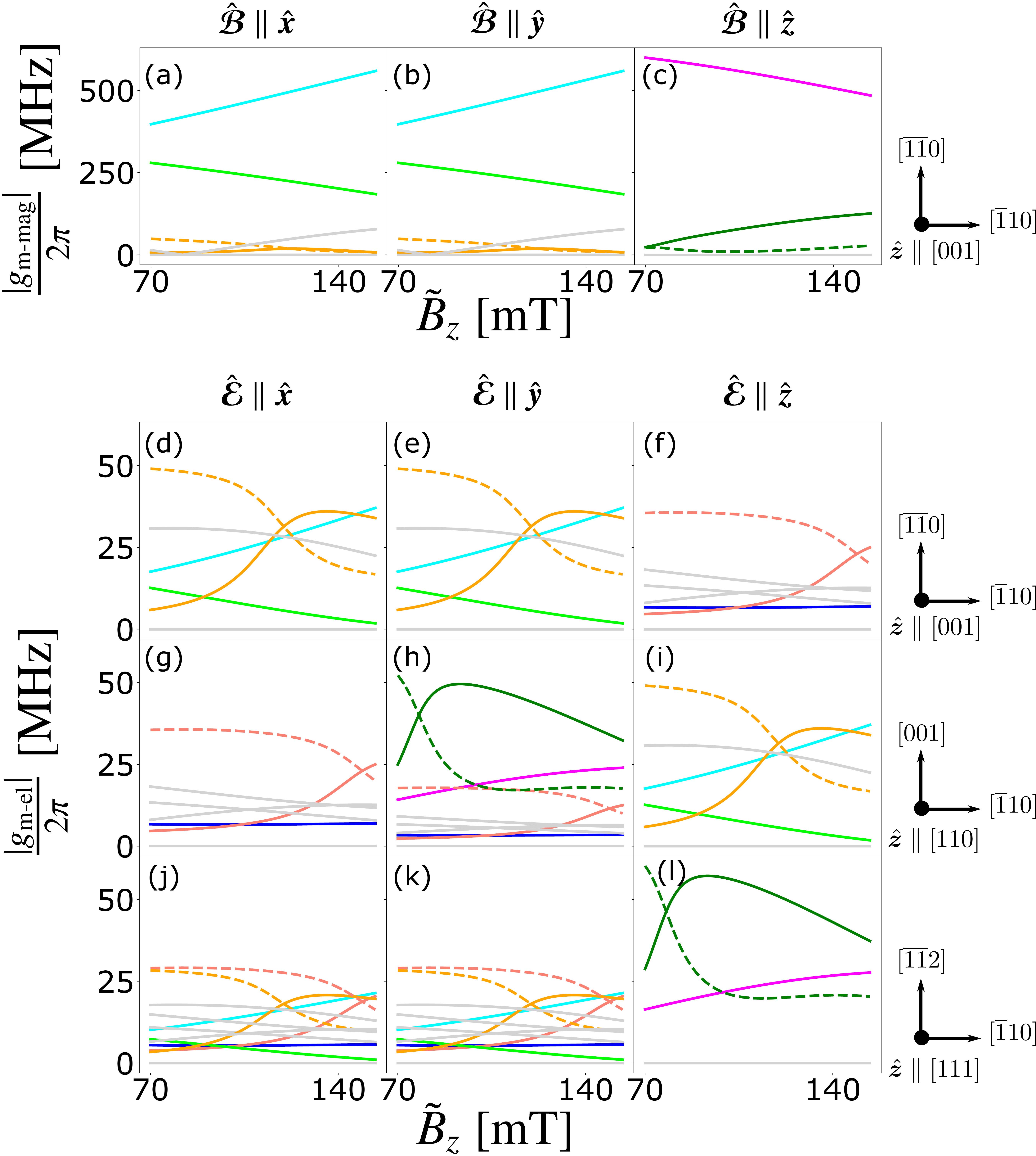}
    \caption{\textbf{Magnetic and electric coupling strengths of magnon modes in the low-temperature SkX.}
    (a-c)~Magnetic coupling constant $|g_{\textrm{m-mag}}^{(n)}|/(2\pi)$ and (d-l)~electric coupling constant $|g_{\textrm{m-el}}^{(n)}|/(2\pi)$ of the lowest 18 energy eigenstates, obtained at $D/J=0.5$, $\Omega=\tilde{\mathsf{E}}_{n}/\hbar$, and $\eta=0.1$ with Eqs.~\eqref{eq:SM_g_mag} and~\eqref{eq:SM_g_el}.
    The rescaled energy eigenvalues of the $n$th magnon mode $\tilde{\mathsf{E}}_{n}$ and magnetic field~$\tilde{B}_z$ are defined in Eqs.~\eqref{eq:SM_tildeEn} and~\eqref{eq:SM_tildeBz}, respectively. For typical parameters of $V=300$~mm$^3$~\cite{Abdurakhimov2019} and $\Omega=4$~GHz for the CCW mode, the maximum amplitudes of magnetic and electric fields are $\CalB_0=2.4$~pT and $\CalE_0=0.7$~mV~m$^{-1}$. 
    In each row, the directions of electromagnetic fields $\hat{\vec{\CalB}}$ and $\hat{\vec{\CalE}}$ are aligned parallel to $\hat{\vec{x}}$, $\hat{\vec{y}}$, and $\hat{\vec{z}}$ axes (from left to right), where $\hat{\vec{z}}$ is parallel to the applied magnetic field and the corresponding crystalline axes are indicated to the right. The color of lines is the same as Fig.~\ref{fig:SkX_band}(b). 
    }
    \label{fig:SM_gfactor}
\end{figure}

We assume the frequency of the cavity mode to be at resonance with the $n$th magnon mode, $\Omega=\tilde{\mathsf{E}}_{n}/\hbar$, and fix the value of $\eta$ as $\eta=0.1$. In this ideal condition, we compute the magnetic and electric coupling strengths of the lowest 18 energy eigenstates in the low-temperature SkX using Eqs.~\eqref{eq:SM_g_mag} and~\eqref{eq:SM_g_el}.
Since the energy eigenvalues are increased by a larger effective coupling constant $\tilde{J}$ compared to the high-temperature SkX, the magnetic coupling strength reaches 500~MHz in the CCW and breathing modes as shown in Fig.~\ref{fig:SM_gfactor}(a-c).
Noting that the damping rates for CCW and breathing modes were estimated approximately as $300$~MHz for CCW and breathing modes in the high-temperature SkX phase~\cite{khanCouplingMicrowavePhotons2021, liensbergerTunableCooperativityCoupled2021a}, the strong coupling regime can be realized in the low-temperature SkX phase. 

The electric coupling strength can also exceed 50~MHz but it is approximately 10 times smaller than the magnetic coupling strength similar to what we extracted for the ferromagnetic phase in Eq.~\eqref{eq:ratio} of the main text. 
As shown in Fig.~\ref{fig:SM_gfactor}(d-l), many magnon modes exhibit a nonzero electric coupling. Also, the coupling strength depends on directions of both the external magnetic field~$\hat{\vec{z}}$ and the cavity mode~$\hat{\vec{\CalE}}$, resulting in a strongly anisotropic MP coupling. Magnon modes with a large electric coupling include the CCW~(solid cyan), breathing~(solid magenta), second elliptic~(solid grey), lowest pentagon~(solid orange), second CCW~(dashed orange), lowest hexgon~(solid green), and second breathing~(dashed green). For experimental observations, we are interested in the lowest-energy elliptic mode~(solid blue), whose coupling constant is $|g_\textrm{m-el}^{(n)}|/(2\pi)\approx 6.8$~MHz when $\hat{\vec{z}},\hat{\vec{\CalE}}\parallel [001]$~[see Fig.~\ref{fig:SM_gfactor}(f)].
In addition, polygon modes with $\ell\ge 4$ show a strong magnetic field dependence in $g_\textrm{m-el}^{(n)}$.  This is a result of hybridization with electrically active modes as discussed in Appendix~\ref{sec: spinwave_skx}. For example, the pentagon mode~(solid orange) shows a large enhancement of $g_\textrm{m-el}^{(n)}$ above $\tilde{B}_z\approx 100$~mT when $\hat{\vec{z}}\parallel [001]$ and $\hat{\vec{\CalE}}\parallel [\overline{1}10]$, as it hybridizes with the second CCW mode~(dashed orange)~[see Fig.~\ref{fig:SM_gfactor}(d)].

\begin{figure}[t]
    \centering
    \includegraphics[width=\columnwidth]{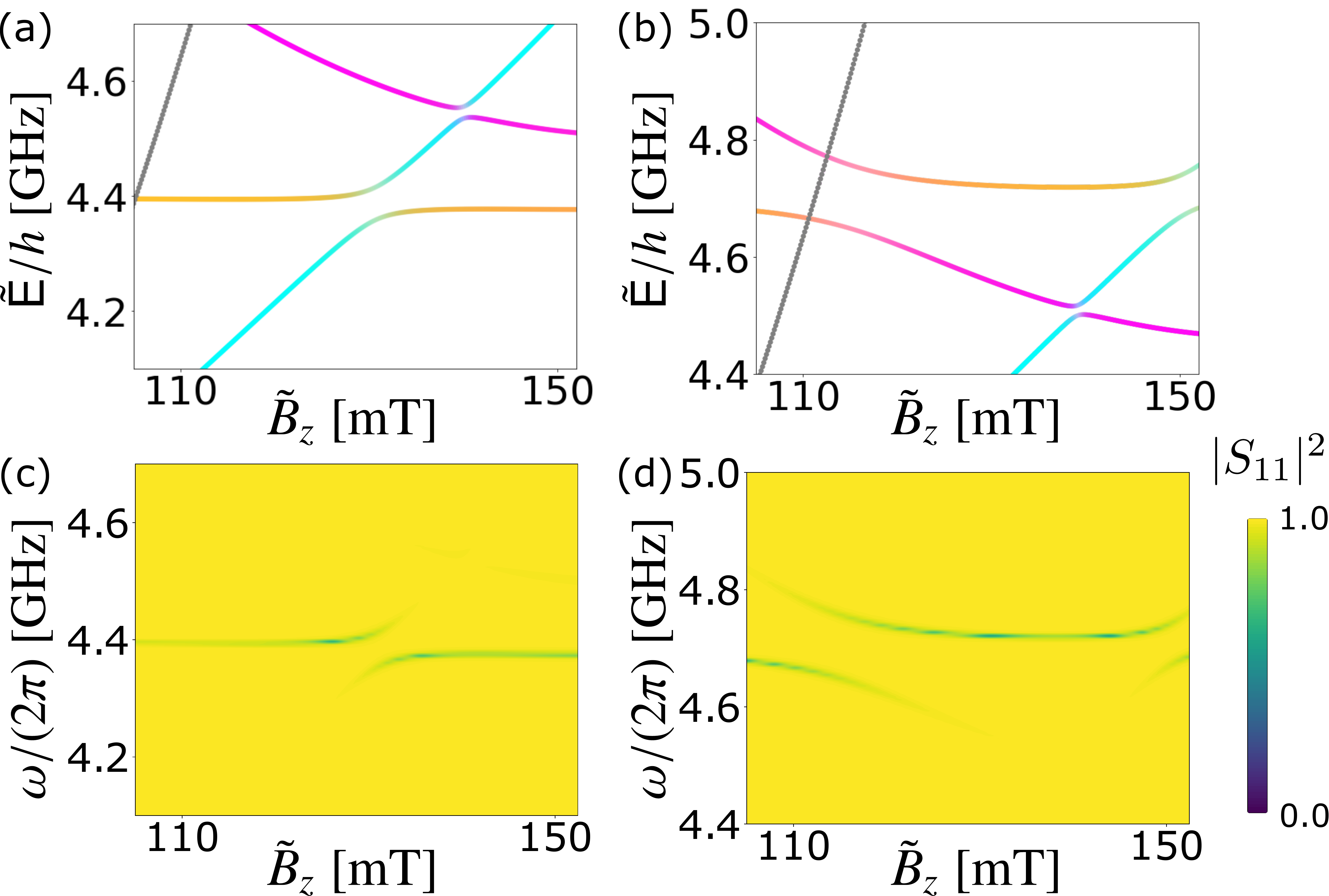}
    \caption{\textbf{Cavity-induced magnon-magnon interaction at off-resonance microwave frequencies.}
    (a,b)~Magnetic field dependence of magnon-photon spectrum of the low temperature SkX at $\vec{k}=0$, where $D/J=0.5$, $\eta=0.1$, $\kappa_a/(2\pi)=1$~MHz, and $\kappa_n/(2\pi)=20$~MHz. The energy eigenvalues~$\tilde{\mathsf{E}}$ and magnetic fields~$\tilde{B}_z$ are rescaled for Cu$_2$OSeO$_3$.
    The color code indicates the probability density of the cavity mode~(dark yellow) and the breathing~(magenta), and CCW~(cyan) modes. Planck's constant is denoted as $h$. (c,d)~Calculated microwave reflection $|S_{11}(\omega)|^2$ using Eq.~\eqref{eq: microwave_S11}.
    In all panels, electromagnetic fields of the cavity modes are directed in $\hat{\vec{\CalB}}\parallel [001]$ and $\hat{\vec{\CalE}}\parallel [\overline{1}10]$ with (a,c)~$\Omega/(2\pi)=4.4$~GHz and (b,d)~$\Omega/(2\pi)=4.7$~GHz. The direction of static magnetic fields is fixed as $\hat{\vec{z}}\parallel[001]$. The SkX is assumed to be placed slightly away from the antinode of the electric component of the cavity mode with $|\vec{\CalB}|\approx 0.1\CalB_0$ and $|\vec{\CalE}|\approx 0.995\CalE_0$~[see Eq.~\eqref{eq: SM_gvalue2}].
    }
    \label{fig:SM_CCW_Br}
\end{figure}

\section{Experimental signature of magnon-photon coupling in skyrmion crystals}
\label{sec:experiment_SkX}
In this section, we discuss experimental signatures of magnon-photon coupling of skyrmion crystals employing the input-output formalism~\cite{Abdurakhimov2019,walls2008quantum}. We consider the following magnon-photon Hamiltonian:
\begin{align}
    \mathcal{H}&=\hbar\left(\Omega-\mathrm{i}\frac{\kappa_a}{2}\right) a^\dagger a+ \mathrm{i} \hbar \sqrt{\kappa_a^\textrm{ext}}a^\dagger a_\textrm{in}-\mathrm{i}\hbar \sqrt{\kappa_a^\textrm{ext}}a_\textrm{out}^\dagger a
    \nonumber\\
    &+\sum_{n=1}^{18} \Big[\left(\tilde{\mathsf{E}}_n-\mathrm{i}\hbar\frac{\kappa_n}{2}\right) b_{0,n}^\dagger b_{0,n}+\hbar g^{(n)}a^\dagger b_{0,n}+\textrm{H.c.}\Big],
    \label{eq:SM_phenomenological}
\end{align}
where $a_\textrm{in/out}$ are introduced as external input/output fields to the cavity, $\kappa_a^\textrm{ext}/(2\pi)=1$~MHz is the coupling constant between microwave cavity modes and external fields, $\Omega$ is the frequency of cavity mode, $g^{(n)}$ is the coupling constant of the $n$th magnon mode given as Eq.~\eqref{eq:SM_gvalue}, and $\kappa_{a}/(2\pi)=1$~MHz is the damping rate of cavity mode~\cite{LachanceQuirion2019}. For the damping rate $\kappa_n$ of the $n\textrm{th}$ magnon mode, we assume $\kappa_n/(2\pi)=20$~MHz for the low-temperature SkX phase~\cite{aqeelMicrowaveSpectroscopyLowTemperature2021b}, which could be realized owing to a very low Gilbert damping constant of $\alpha\approx 10^{-4} $ at 5~K~\cite{stasinopoulosLowSpinWave2017}. 
Since anomalous terms such as $a^\dagger b_{0,n}^\dagger$ are neglected in the rotating wave approximation, Eq.~\eqref{eq:SM_phenomenological} can be diagnoalized by a unitary matrix~$U$:
\begin{align}
\mathcal{H}&=\hbar\left(\Omega'-\mathrm{i}\frac{\kappa_a'}{2}\right)a'^\dagger a'+\sum_{n=1}^{18}\left(\tilde{\mathsf{E}}_n'-\mathrm{i}\hbar \frac{\kappa_n'}{2}\right)b_{0,n}'^\dagger b'_{0,n}
\nonumber\\
&+\mathrm{i}\hbar \sqrt{\kappa_a^\textrm{ext}}\Psi_j^{\prime \dagger}
U_{j0} a_\textrm{in}-\mathrm{i}\hbar \sqrt{\kappa_a^\textrm{ext}}a_\textrm{out}^\dagger U_{0j}^\dagger \Psi'_j,
\end{align}
with $\Psi_j=(a,b_{0,1},\cdots, b_{0,18})^\text{T}$, $\Psi'_j=(a',b_{0,1}',\cdots, b_{0,18}')^\text{T}$, and $\Psi'_j=U_{ij}\Psi_j$ for $i,j=0,1,\cdots, 18$.

The expectation value of each operator in Eq.~\eqref{eq:SM_phenomenological} evolves as $\frac{\partial \braket{o}}{\partial t}=-\frac{\mathrm{i}\braket{[o,H]}}{\hbar}$, where $\braket{o}$ represents the expectation value of an operator $o$. Assuming that $a(t)=a\mathrm{e}^{-\mathrm{i}\omega t}$ and $b_{0,n}(t)=b_{0,n}\mathrm{e}^{-\mathrm{i}\omega t}$, the equations of motion are obtained as~\cite{Abdurakhimov2019}
\begin{subequations}
\begin{gather}
-\mathrm{i}\omega a=
    \left(-\mathrm{i}\Omega-\frac{\kappa_a}{2}\right)a-\sum_{n=1}^{18} \mathrm{i} g^{(n)}b_{0,n}+\sqrt{\kappa_a^\textrm{ext}}a_\textrm{in},
    \\
-\mathrm{i}\omega b_{0,n}=
    -\mathrm{i}g^{(n)} a+\left(-\mathrm{i}\frac{\tilde{\mathsf{E}}_n}{\hbar}-\frac{\kappa_n}{2}\right) b_{0,n},
    \\
a_\textrm{in}+a_\textrm{out}=
    \sqrt{\kappa_a^\textrm{ext}}a,
\end{gather}
\end{subequations}
where the third equation is introduced as a boundary condition between the cavity and external fields~\cite{walls2008quantum}. Using the diagonalized basis, we obtain the following equations:
\begin{subequations}
\begin{gather}
-\mathrm{i}\omega a'=
    \left(-\mathrm{i}\Omega'-\frac{\kappa'_a}{2}\right)a'+U_{00}\sqrt{\kappa_a^\textrm{ext}}a_\textrm{in},
\\
-\mathrm{i}\omega b_{0,n}'=
    \left(-\mathrm{i}\tilde{\mathsf{E}}'_n/\hbar-\frac{\kappa'_n}{2}\right)b_{0,n}'+U_{n0}\sqrt{\kappa_a^\textrm{ext}}a_\textrm{in},
\\
a_\textrm{in}+a_\textrm{out}=
    \sqrt{\kappa_a^\textrm{ex}}(U^\dagger_{00}a'+U^\dagger_{0n}b'_{0,n}).
\end{gather}
\end{subequations}
Finally, it is straightforward to derive the expression for microwave reflection $|S_{11}(\omega)|^2=|a_\textrm{out}/a_\textrm{in}|^2$ as
\begin{align}
|&S_{11}(\omega)|^2= \nonumber \\
&\Big|-1+\kappa_a^\textrm{ext}\Big(\frac{|U_{00}|^2}{\mathrm{i}(\Omega'-\omega)+\kappa_a'/2}+\sum_{n=1}^{18} \frac{|U_{n0}|^2}{\mathrm{i}(\tilde{\mathsf{E}}_n'/\hbar-\omega)+\kappa_n'/2}\Big)\Big|^2.
\label{eq: microwave_S11}
\end{align}
When $g^{(n)}=0$, the input field is reflected almost perfectly except at $\omega=\Omega$ with a line width given by $\kappa_a$ and $\kappa_a^\textrm{ext}$. In contrast, the MP coupling results in the scattering of photons when they are at resonance. Hence, magnon-photon hybrid systems can be studied by measuring $|S_{11}(\omega)|^2$ with changing the frequency~$\omega$ of the input field.

\begin{figure}[t]
    \centering
    \includegraphics[width=\columnwidth]{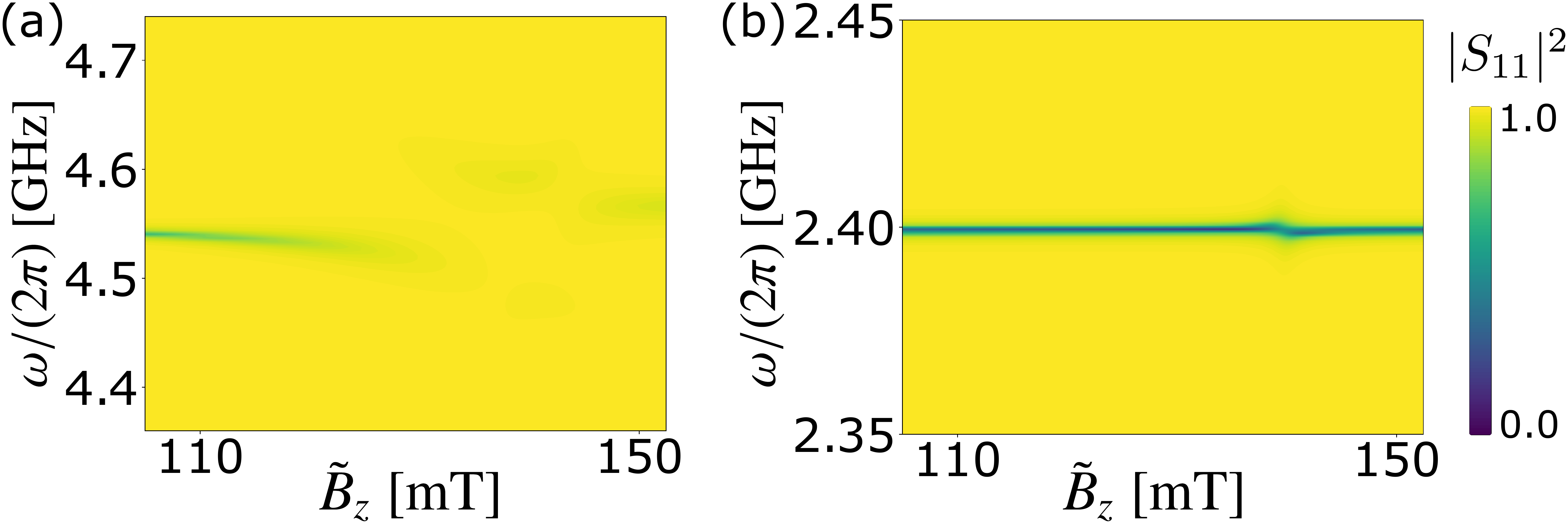}
    \caption{
    \textbf{Cavity-induced magnon-magnon interaction and MP hybridization of the magnetically dark elliptic mode at a larger damping rate.}
    Calculated microwave reflection $|S_{11}(\omega)|^2$ using Eq.~\eqref{eq: microwave_S11}, where $D/J=0.5$, $\eta=0.1$, $\kappa_a/(2\pi)=1$~MHz, and $\kappa_n/(2\pi)=100$~MHz. The energy eigenvalues~$\tilde{\mathsf{E}}$ and magnetic fields~$\tilde{B}_z$ are rescaled for Cu$_2$OSeO$_3$. Electromagnetic fields of the cavity modes are directed in (a)~$\hat{\vec{\CalB}}\parallel [001]$ and $\hat{\vec{\CalE}}\parallel [\overline{1}10]$ at $\Omega/(2\pi)=4.55$~GHz and (b)~$\hat{\vec{\CalB}}\parallel [\overline{1}10]$ and $\hat{\vec{\CalE}}\parallel [001]$ at $\Omega/(2\pi)=2.4$~GHz. The direction of static magnetic fields is fixed as $\hat{\vec{z}}\parallel[001]$. We choose the same range of magnetic fields and frequencies as Fig.~\ref{fig: S11_plot}(b) and~(d) of the main text for (a) and (b), respectively. The SkX is assumed to be placed slightly away from the antinode of the electric component of the cavity mode with $|\vec{\CalB}|\approx 0.1\CalB_0$ and $|\vec{\CalE}|\approx 0.995\CalE_0$~[see Eq.~\eqref{eq: SM_gvalue2}].
    }
    \label{fig:SM_kappa50}
\end{figure}

In the following, we assume that the SkX is placed slightly away from the antinode of the electric component of the cavity mode with $\hat{\vec{z}}\parallel[001]$. Noting that the total energy density of electromagnetic fields $(|\vec{\CalB}|^2/\mu_0+\epsilon_0|\vec{\CalE}|^2)/2$ is constant inside the microwave cavity, we take $|\vec{\CalB}|=0.1\CalB_0$ and $|\vec{\CalE}|=0.995\CalE_0$. The MP coupling is then written as 
\begin{equation}
    \hbar g^{(n)}=-\sqrt{\frac{\hbar\Omega\mu_0 \rho_s\eta}{2}}\,(0.1\hat{\vec{\CalB}}\cdot \vec{\CalM}^{(n)}+0.995c\hat{\vec{\CalE}}\cdot \vec{\CalP}^{(n)}),\label{eq: SM_gvalue2}
\end{equation}
with the speed of light $c$ and $\rho_s= 5.67\times 10^{27}$~m$^{-3}$. We set $\eta=0.1$. Since the magnetic coupling is stronger than the electric coupling approximately by a factor of eleven~[see Eq.~\eqref{eq:ratio} of the main text], this setup compensates for the difference in their coupling strengths.

First, we consider the case when the cavity mode is at resonance with the CCW and breathing modes. 
Choosing the frequency of the cavity mode at $\Omega/(2\pi)=4.55$~GHz, we have $|\vec{\CalB}|\approx 0.25$~pT and $|\vec{\CalE}|\approx 0.75$~mV~m$^{-1}$ for $V=300$~mm$^{-3}$.
Importantly, both modes are magnetically and electrically active with large coupling constants as shown in Fig.~\ref{fig:SM_gfactor}. By applying the cavity mode with $\hat{\vec{\CalB}}\parallel [001]$ and $\hat{\vec{\CalE}}\parallel [\overline{1}10]$, it is coupled magnetically and electrically with the breathing and CCW modes, respectively. This allows cavity-induced magnon-magnon interactions. 
As demonstrated in Fig.~\ref{fig: S11_plot}(a) and~(b) of the main text, the cavity-induced magnon-magnon interaction is the most prominent at the triple resonance point among the cavity, CCW, and breathing modes. However, even if the cavity mode frequency is slightly detuned from the triple resonance frequency, we can realize the magnon-magnon interaction via virtual processes mediated by the cavity mode. As shown in Fig.~\ref{fig:SM_CCW_Br}(a) and (b), we find small hybridization gaps between the CCW and breathing modes with the blue-detuned or red-detuned cavity mode. Unlike in the triple resonance condition, the anticrossing between the CCW and breathing modes are not directly observed in the microwave reflection $|S_{11}(\omega)|^2$~[see Fig.~\ref{fig:SM_CCW_Br}(c) and (d)].
We also discuss the dependence on the damping rate $\kappa_n$ of magnon modes. When $\kappa_n<|g_\textrm{m-el}^\textrm{CCW}|$, it is in the strong coupling regime with clear signatures of the level anticrossings as shown in Fig.~\ref{fig: S11_plot}(a) of the main text. In contrast, when $\kappa_n>|g_\textrm{m-el}^\textrm{CCW}|$, the hybridization gap becomes less pronounced~\cite{khanCouplingMicrowavePhotons2021}. However, we can still see that $|S_{11}(\omega)|^2$ is increased significantly when the CCW and breathing modes intersect with the cavity mode, as shown in Fig.~\ref{fig:SM_kappa50}(a).

Second, we consider a purely electric coupling between the cavity mode and lowest-energy elliptic mode. Electromagnetic fields of the cavity mode are directed in $\hat{\vec{\CalB}}\parallel [\overline{1}10]$ and  $\hat{\vec{\CalE}}\parallel [001]$ at $\Omega/(2\pi)=2.4$~GHz. Although the coupling constant is not as large as that of the third elliptic mode~[see Fig.~\ref{fig:SM_gfactor}(f)], the damping rate is expected to be much smaller in the lowest-energy elliptic mode. Furthermore, the cavity mode is only coupled to the elliptic mode at this frequency, allowing clear experimental signatures. Assuming $\kappa_n/(2\pi)=20$~MHz, we obtain a small level repulsion between the cavity and elliptic modes as shown in Fig.~\ref{fig: S11_plot}(d) of the main text. As we increase the damping rate $\kappa_n/(2\pi)$ to 100~MHz, an anticrossing between them is no longer observed, as shown in Fig.~\ref{fig:SM_kappa50}(b); however, a very sharp increase in $|S_{11}(\omega)|^2$ is still observed with a small frequency shift of the cavity mode where these two modes intersect.

\section{Quantum gates mediated by magnons and photons}
\label{sec:unitary_gate}

In Sec.~\ref{sec: QI_application}, we demonstrate magnon-mediated SWAP and SPLIT operations and photon-mediated SPLIT operations using a semiclassical approach. 
Here, we discuss the connection between our results and two-qubit quantum gates, following Ref.~\onlinecite{trevillian2020unitary,*trevillian2021unitary}. 
The standard basis for quantum gates is called the computational basis, which labels $\ket{0}_{1q}$ and $\ket{1}_{1q}$ as the ground state and excited state of a single qubit, respectively. Since a qubit is represented as a superposition of $\ket{0}_{1q}$ and $\ket{1}_{1q}$, a single-qubit quantum gate is represented by a $2\times 2$ unitary matrix. Similarly, a two-qubit quantum gate is represented as a $4\times 4$ unitary matrix $U$ with elements $U_{ij} = \braket{i|U_A|j}$, where $i$ and $j$ run over the basis states $0$ to $3$, which read $|0\rangle= |00\rangle$, $|1\rangle = |10\rangle$, $|2\rangle = |01\rangle$, and $|3\rangle = |11\rangle$.

Below, we consider magnon-mediated photon gates (as in Sec.~\ref{sec: QI_application1}), although the following argument is also applicable to photon-mediated magnon gates (as in Sec.~\ref{sec:QI_application2}).
Assuming that two photon modes $a_1$ and $a_2$ are strongly coupled to $\ket{1}$ and $\ket{2}$, respectively, quantum operations mediated by magnons $b$ lead to gate operations between $\ket{1}$ and $\ket{2}$.
Hence, they are represented as
\begin{equation}
    U=
    \begin{pmatrix}
      1&0&0&0\\
      0&U'_{11}&U'_{12}&0\\
      0&U'_{21}&U'_{22}&0\\
      0&0&0&1
    \end{pmatrix},
\end{equation}
where 
\begin{equation}
    U'=
    \begin{pmatrix}
      U'_{11}&U'_{12}\\
      U'_{21}&U'_{22}
    \end{pmatrix}
\end{equation}
is a $2\times 2$ unitary matrix, describing the quantum operation mediated by magnons. 
To derive the expression of $U'$, we rely on the semiclassical photon-magnon coupled model in Sec.~\ref{sec: QI_application1} and use a transfer matrix $T$ to relate initial and final quantum states~\cite{trevillian2020unitary,*trevillian2021unitary}:
\begin{equation}
	\begin{pmatrix}
	a_1 \\ a_2 \\ b
	\end{pmatrix}_\text{final}
    =
    T
	\begin{pmatrix}
	a_1 \\ a_2 \\ b
	\end{pmatrix}_\text{initial}.
\end{equation}
We recall that $a_1$ and $a_2$ are the complex amplitudes of two photon modes and $b$ that of a magnon mode. 
Crucially, using a suitable field protocol, the population of magnons does not change after mediating the quantum operation. Thus, we can write the transfer matrix 
in a block diagonal form as~\cite{trevillian2020unitary,*trevillian2021unitary}
\begin{equation}
	T=
	\begin{pmatrix}
	U'_{11}&U'_{12}&0\\
    U'_{21}&U'_{22}&0\\
    0&0&1
	\end{pmatrix},
\end{equation} 
where the effect of decoherence is neglected. 
If we assume that the effective coupling between photons can be written as $H'=G(t)\sigma_x$ with a phenomenological coupling strength $G(t)$, we can write $U'$ as $U'=\mathrm{e}^{\mathrm{i}\phi}(\alpha\sigma_0+\mathrm{i}\beta\sigma_x)$, with $\alpha$ and $\beta$ being real, and $\sigma_i$ denoting the identity matrix and Pauli matrices for $i=0,x,y,z$. 

As shown in Sec.~\ref{sec: QI_application1}, the magnon-mediated SWAP operation results in quantum transduction between $\ket{1}$ and $\ket{2}$.
This is realized when $\alpha=0$ and $\beta=1$. We note that the complex phase of $\phi$ depends on the choice of field protocols. In particular, we have $U'_\textrm{SWAP}=\sigma_x$ when $\phi=-\pi/2$. 
Similarly, the magnon-mediated SPLIT operation results in splitting of $\ket{1}$ and $\ket{2}$ into a superposition of $\ket{1}$ and $\ket{2}$.
Thus, the SPLIT gate is realized at $\alpha=\beta=1/\sqrt{2}$, where it is represented as $U'_\textrm{SWAP}=e^{i\phi}(\sigma_0+i\sigma_x)/\sqrt{2}$. We note that it is equivalent to a square root of SWAP gate when $\phi=\pi/4$, which is entangling and thus universal~\cite{lossQuantumComputationQuantum1998}.
Since SPLIT gates also generate entanglement between qubits, a universal set of quantum gates can be constructed with magnon-mediated quantum operations~\cite{bremnerPracticalSchemeQuantum2002}.
\\

\section{Magnon Stark effect in the collinear phase of {Cu$_2$OSeO$_3$}}
\label{sec:p2}

In the expansion of the microscopic electric dipole moments in terms of magnon variables, 
Eq.~\eqref{eq:expansion-of-p},
the electric dipole moment expectation value of magnons is encoded in the bilinear $\vec{p}_i^{(2)}$. For Cu$_2$OSeO$_3$, it reads
\begin{widetext}
\begin{align}
	\vec{p}_i^{(2)} =
	\frac{\lambda S}{2}
	\begin{pmatrix}
		- ( b_i b_i + 6 b_i^\dagger b_i + b_i^\dagger b_i^\dagger ) \sin \theta_i \cos \theta_i \sin \phi_i
		+ \mathrm{i} ( b_i - b_i^\dagger ) (b_i + b_i^\dagger) \sin \theta_i \cos \phi_i
		\\
		- ( b_i b_i + 6 b_i^\dagger b_i + b_i^\dagger b_i^\dagger ) \sin \theta_i \cos \theta_i \cos \phi_i
		- \mathrm{i} ( b_i - b_i^\dagger ) (b_i + b_i^\dagger) \sin \theta_i  \sin \phi_i
		\\
		\frac{1}{4} \left( 
			3 (b_i - b_i^\dagger)^2 \sin (2 \phi_i) + (b_i b_i + 6 b_i^\dagger b_i + b_i^\dagger b_i^\dagger) \cos(2\theta_i) \sin(2\phi_i)
			- 4 \mathrm{i} (b_i - b_i^\dagger)(b_i + b_i^\dagger) \cos \theta_i \cos(2\phi_i)
		\right)
	\end{pmatrix}
	\label{eq:quadratic-polarization}
\end{align}
\end{widetext}
and contains both normal magnon coupling ($b_i^\dagger b_i$) and anomalous coupling ($b_i b_i$). In the ferromagnetic phase, we may drop the index $i$ and the anomalous coupling, leaving us with
$
    \vec{p}^{(2)} = - \frac{3}{S} b^\dagger b \vec{p}^{(0)},
$
where $\vec{p}^{(0)}$ is the ground state polarization given in Eq.~\eqref{eq:GS-polarization}. Consequently, $\vec{p}^{(2)}$ can only be nonzero if $\vec{p}^{(0)}$ is nonzero.

\begin{figure}[t]
    \centering
    \includegraphics[width=0.8\columnwidth]{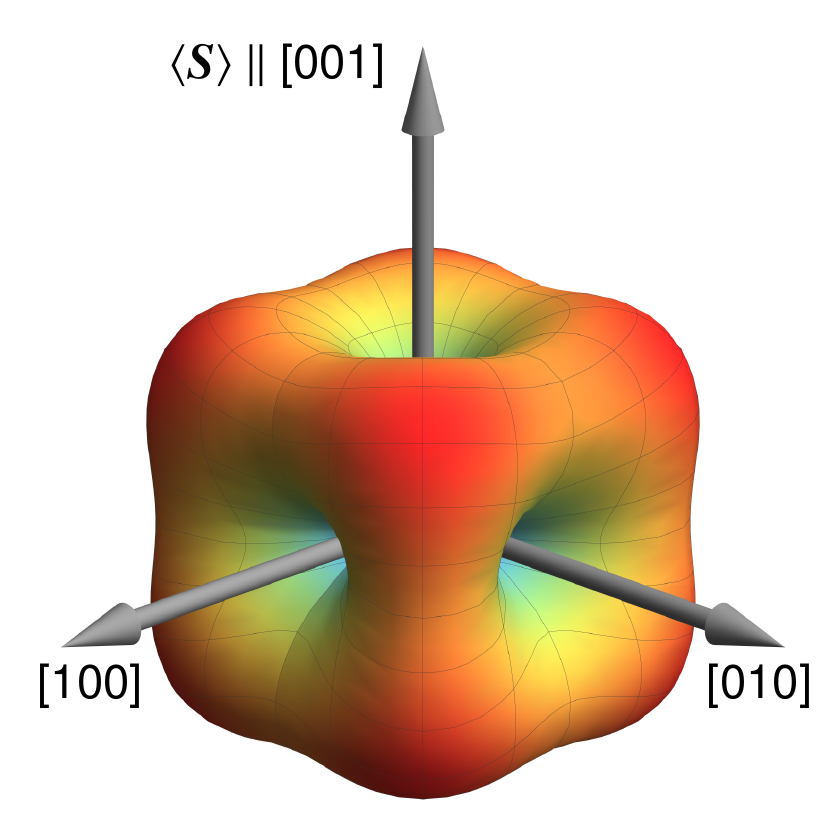}
    \caption{\textbf{Magnitude of the electric ground state polarization $\vec{p}^{(0)}$ and the bilinear $\vec{p}^{(2)}$ as a function of the ground state spin polarization $\langle \vec{S} \rangle$ in the collinear phase of Cu$_2$OSeO$_3$.}
    For $\langle \vec{S} \rangle$ along the $a$, $b$, or $c$ directions, which respectively correspond to $[100]$, $[010]$, and $[001]$, there is zero $\vec{p}^{(0)}$ and $\vec{p}^{(2)}$. Maximal $\vec{p}^{(0)}$ and $\vec{p}^{(2)}$ are found for $\langle \vec{S} \rangle$ along the $\langle 111 \rangle$ directions.
    }
    \label{fig:P0}
\end{figure}

\begin{figure}[t]
    \centering
    \includegraphics[width=\columnwidth]{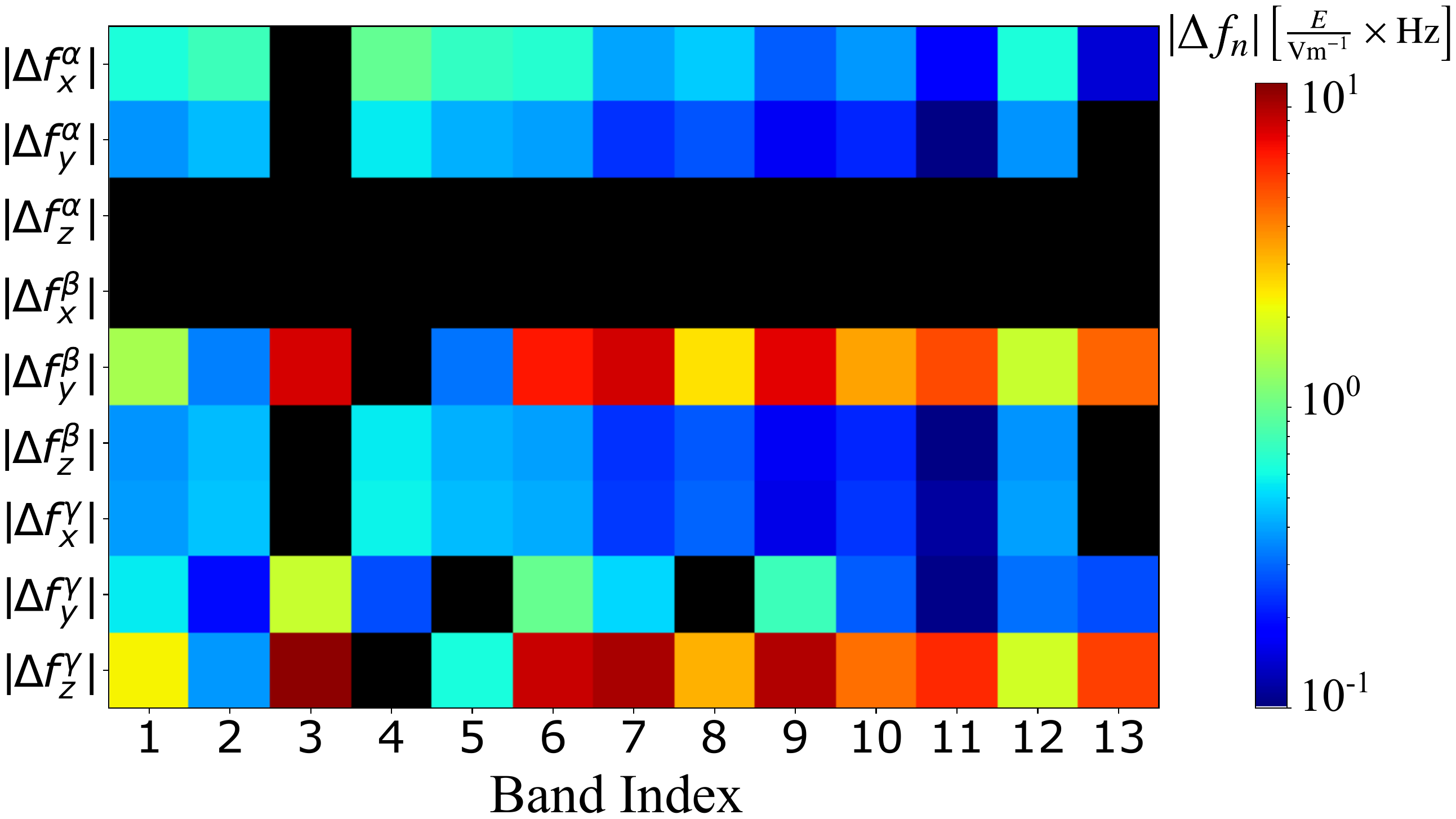}
    \caption{
    \textbf{Magnon Stark effect of low-energy magnon modes in SkXs.} The amplitude of frequency shift~$\Delta f_n$ is plotted on a logarithmic scale, obtained at $D/J=0.5$ and $\mathsf{g} \mu_B B_z J / (D^2S)=0.8$~($\tilde{B}_z=\unit[112]{mT}$). The superscripts $\alpha,\,\beta, \textrm{ and}~\gamma$ denote $\hat{\vec{z}}\parallel [001]$, $[110]$, and $[111]$, respectively, and the subscript denotes the direction of static electric field $\hat{\boldsymbol{E}}$. For all cases, the $x$-axis is taken along $[\overline{1}10]$. Black regions indicate the frequency shift smaller than $10^{-1}$. See Table~\ref{table: Multipole_expansion} for the description of magnon modes.
    }
    \label{fig:SM_SkXStark}
\end{figure}

Figure \ref{fig:P0} depicts the magnitude of $\vec{p}^{(0)}$ and $\vec{p}^{(2)}$ upon variation of the direction of $\langle \vec{S} \rangle$. As indicated by the red lobes, maximal $\vec{p}^{(2)}$ is found for $\langle \vec{S} \rangle \parallel [111]$ (along the $abc$ direction; $\theta = \arctan \sqrt{2}$, $\phi = \pi/4$), for which we find $\vec{p}^{(2)}_{[111]} = - \lambda b^\dagger b (1,1,1)^\text{T}$ (recall $S=1$). Thus, applying a static electric field $\vec{E}$ parallel to $\vec{p}^{(0)} \parallel [111]$ causes a Stark frequency shift of magnons of 
\begin{align}
    \Delta f
    = 
    \unit[14.7]{Hz} \times \frac{E}{\unit{V/m}}
    \quad
    \text{for}
    \quad
    \langle \vec{S} \rangle, \vec{E} \parallel [111].
\end{align}
For a field strength between $E = \unit[10^6]{V/m}$ and $\unit[10^8]{V/m}$, the resulting magnonic Stark shift is estimated as $ \Delta f = \unit[14.7]{MHz}$ and $\unit[1.47]{GHz}$.

\section{Magnon Stark effect in skyrmion crystals}
\label{sec: SkX_Stark}
The bilinear term $\vec{p}^{(2)}_i$ contains both normal and anomalous terms as shown in Appendix~\ref{sec:p2}. While the anomalous term ($b^\dagger_i b_i^\dagger$) vanishes in the magnon Stark effect of collinear ferromagnets, it becomes finite in noncollinear magnets. From Eq.~\eqref{eq:quadratic-polarization}, we have
\begin{equation}
    \vec{p}_i^{(2)}=-\frac{3}{S}\,\vec{p}_i^{(0)}b^\dagger_ib_i+\vec{\eta}_ib_ib_i+\vec{\eta}_i^* b^\dagger_ib^\dagger_i,
\end{equation}
with 
\begin{align}
\vec{\eta}_i &=
\frac{\lambda S}{2}
\begin{pmatrix}
(\mathrm{i}\cos\phi_i-\cos\theta_i\sin\phi_i)\sin\theta_i
\\
-(\cos\phi_i\cos\theta_i+\mathrm{i}\sin \phi_i)\sin\theta_i
\\
\frac{1}{4}(3+\cos2\theta_i)\sin2\phi_i-\mathrm{i}\cos2\phi_i\cos\theta_i\sin2\phi_i
\end{pmatrix}.
\end{align}
Assuming the spatially uniform and static electric field $\vec{E}(\vec{r}_i,t)=\vec{E}$, we perform the Fourier transform to obtain 
\begin{equation}
    H
    =
    -\sum_{\vec{k}} \sum_{j=1}^{N_\textrm{b}}\vec{E}\cdot\left(-\frac{3}{S}\, \vec{p}_j^{(0)}b^\dagger_{\vec{k},j} b_{\vec{k},j}+\vec{\eta}_jb_{\vec{k},j}b_{-\vec{k},j}+\vec{\eta}_j^*b^\dagger_{\vec{k},j}b^\dagger_{-\vec{k},j}\right).
\end{equation}
From $(b_{\vec{k},j},b^{\dagger}_{-\vec{k},j})^T=\sum_{n=1}^{N_\textrm{b}}T^{j,n}_{\vec{k}}(b_{\vec{k},n},b^{\dagger}_{-\vec{k},n})^T$, the frequency shift of $n$-th magnon mode is given as
\begin{align}
\Delta f_{n}&=\frac{1}{h}\hat{\vec{E}}\cdot \sum_j\Big\{-\frac{3}{S}\, \vec{p}_j^{(0)}\left(|u_{\vec{k}}^{j,n}|^2+|v_{\vec{k}}^{j,n}|^2 \right)+2\vec{\eta}_j u_{\vec{k}}^{j,n}\left(v_{\vec{k}}^{j,n}\right)^*\nonumber\\
&+2\vec{\eta}_j^*\left(u_{\vec{k}}^{j,n}\right)^*v_{\vec{k}}^{j,n}\Big\} \textrm{Hz} \times \frac{E}{\textrm{V/m}},
\end{align}
where $\hat{\vec{E}}$ is a unit vector parallel to the electric field.

Figure~\ref{fig:SM_SkXStark} shows the amplitude of $\Delta f_{n}$ for the low-energy magnon modes in SkXs. As shown in the top three rows, the magnon Stark effect does not vanish despite $\sum_i\vec{p}^{(0)}_i=0$ for $\hat{z}\parallel [001]$, where $|\Delta f_{\textrm{\,Breathing}}|/E\approx 0.95~\textrm{Hz}/(\textrm{Vm}^{-1})$. The maximum frequency shift is obtained as $|\Delta f_{\,\textrm{CCW}}|\approx 1.1$~GHz for $\hat{z}\parallel [111]$ and $E=10^8$~V/m. 
Importantly, $\Delta f_{n}$ depends strongly on the magnetization direction~$\hat{\vec{z}}$, the external electric field~$\hat{\vec{E}}$, and the band index~$n$. Thus, the magnon Stark effect enables a selective control of magnon mode frequencies in SkXs.


%

\end{document}